\documentclass{aa}

\usepackage{graphicx}
\usepackage{xcolor}
\usepackage{txfonts}

\usepackage{amsmath,amssymb,siunitx}

\newcommand{\CII}{[C{\scriptsize\,II}]}

\newcommand{\HII}{H{\scriptsize\,II~}}

\let\originallesssim\lesssim
\let\originalgtrsim\gtrsim

\DeclareRobustCommand{\lesssim}{%
  \mathrel{\mathpalette\lowersim\originallesssim}%
}
\DeclareRobustCommand{\gtrsim}{%
  \mathrel{\mathpalette\lowersim\originalgtrsim}%
}

\makeatletter
\newcommand{\lowersim}[2]{%
  \sbox\z@{$#1<$}%
  \raisebox{-\dimexpr\height-\ht\z@}{$\m@th#1#2$}%
}
\makeatother
\makeatletter
\newcommand{\Spvek}[2][r]{%
  \gdef\@VORNE{1}
  \left(\hskip-\arraycolsep%
    \begin{array}{#1}\vekSp@lten{#2}\end{array}%
  \hskip-\arraycolsep\right)}

\def\vekSp@lten#1{\xvekSp@lten#1;vekL@stLine;}
\def\vekL@stLine{vekL@stLine}
\def\xvekSp@lten#1;{\def\temp{#1}%
  \ifx\temp\vekL@stLine
  \else
    \ifnum\@VORNE=1\gdef\@VORNE{0}
    \else\@arraycr\fi%
    #1%
    \expandafter\xvekSp@lten
  \fi}
\makeatother
\begin{document}

   \title{The embedded cluster RCW~108}
   \subtitle{I. An ALMA perspective on feedback-associated compact cores}

   \author{F. Peißker\inst{1}
          \and N. Schneider\inst{1}
          \and S. Kabanovic\inst{1}
          \and S. Dannhauer\inst{1}
          \and \'A. S\'anchez-Monge\inst{2,3}
          \and R. Simon\inst{1}
          \and E. Bordier\inst{1}
          \and L. Labadie\inst{1}
          \and R. Grellmann\inst{1}
          \and C. Nagar\inst{1}
          \and P. Schilke\inst{1}
          \and M. Wiegand\inst{1}
          \and L. Großekathöfer\inst{1}
          \and N. Schmökel\inst{1}
          \and M. Melamed\inst{1,4}
          \and S. Pfalzner\inst{5}
          }

   \institute{I. Physikalisches Institut, Universit\"at zu K\"oln,
             Z\"ulpicherstr. 77, 50937 K\"oln, Germany\\
              \email{peissker@ph1.uni-koeln.de}\and
              Institut de Ciències de l’Espai (ICE), CSIC, Campus UAB, Carrer de Can Magrans s/n, E-08193, Bellaterra, Barcelona, Spain \and
              Institut d’Estudis Espacials de Catalunya (IEEC), E-08860, Castelldefels, Barcelona, Spain \and
              Max-Planck-Institut f\"ur Radioastronomie, Auf dem H\"ugel 69, 53121 Bonn, Germany \and
              Jülich Supercomputing Center, Forschungszentrum Jülich, 52428 Jülich, Germany
             }

   \date{Received ---; accepted ---}

  \abstract
   {}
   {RCW~108 is a young embedded \HII\ region and molecular cloud complex in the Ara~OB1 association. Its central massive stellar population, including the O9-type source IRS~29, makes the region a suitable target for studying the compact dense-gas structures that form in the vicinity of massive stars. In this work, we use ALMA Band~6 data to investigate the compact continuum core population in RCW~108 and to examine whether its spatial distribution is consistent with feedback-associated compression.}
   {We analyze ALMA Band~6 continuum emission at 233.0~GHz ($\sim$1.3 mm) together with molecular-line observations of {C$^{18}$O,} $\rm ^{13}CO$, H$_2$CO, and CH$_3$OH. The compact core sample is defined from the 233.0~GHz continuum emission {using \texttt{astrodendro}. We extract C$^{18}$O\,(2-1) and $^{13}$CO\,(2-1) spectra directly over the continuum-defined core masks and derive velocity dispersions for the virial analysis, propagating the uncertainties of fluxes, dust temperature, dust opacity, gas-to-dust ratio, line widths, and radii with a Monte Carlo approach.}
   In addition, we use the para-H$_2$CO transition lines to determine the kinetic temperatures of embedded cores observed at 1.3 mm. Chandra X-ray observations allow us to locate the stellar population relative to the ALMA continuum cores.}
   {We identify {33} compact cores in RCW~108 based on ALMA continuum observations. The majority of these cores are distributed along arc-like molecular structures around the central embedded stellar population. {For all {14} cores above ${1.0}\,M_\odot$, we detect C$^{18}$O emission.}
   Several cores have low virial parameters and are consistent with being gravitationally bound or prone to collapse under the adopted assumptions. From the 233.0~GHz continuum emission, we estimate a total continuum core mass of $M_{\rm total}={69.4}^{+{29.9}}_{-{21.3}}\,M_{\odot}$. The core mass function {above ${1.0}\,M_\odot$} {has a maximum-likelihood} slope of $\alpha_{\rm CMF} = {1.79}^{+{0.26}}_{-{0.16}}$. Para-H$_2$CO line ratios indicate warm molecular gas, with kinetic temperatures {of $\approx$81-114~K for four cores, all located in the arc-like filaments}.}
   {The spatial coincidence between continuum cores, arc-like molecular gas, and the central massive stellar population is consistent with a scenario in which feedback from IRS 29 and the embedded cluster may have influenced the dense-gas morphology of RCW 108. The presence of cores with low virial parameters $\alpha_{\rm vir}$ and the warm molecular gas, indicative of local heating, are consistent with the possibility of ongoing or future star formation in the compact gas structures. {However, statistical and observational limitations of the core-sample and data do not uniquely support feedback-driven scenarios but also allow for alternative interpretations, such as pre-existing filament structures or spontaneously formed cores.}}


\maketitle

\section{Introduction}
\label{sec:intro}

Filaments and compact molecular structures play an essential role in the star-formation process \citep[see][and references therein for a review]{Pineda2023}. Far-infrared dust observations with {\sl Herschel} and velocity-resolved molecular-line studies have shown that low-mass stars preferentially form along filaments through fragmentation \citep[e.g.][]{Inutsuka1997, Marsh2016, Koenyves2020}. In contrast, massive stars and clusters are often found in dense {hubs} where filamentary structures converge \citep[e.g.][]{Schneider2012, Kirk2013, Kumar2020}. {Within these hubs, clumps of $\sim$0.3-3~pc \citep{Williams2000, Bergin2007} fragment into massive dense cores (MDCs) of $\sim$0.1~pc and a few 10~M$_\odot$, which are considered to be the immediate progenitors of one or more high-mass stars \citep{Motte2007, Beuther2007, Hosokawa2009, Bontemps2010, Motte2018}. At the resolution of ALMA, MDCs are further resolved into compact cores of a few 10$^3$~AU \citep{Louvet2024, Coletta2025}, which we refer to as cores in this work \citep[see also][]{Andre2014}. Whether these cores collect their mass before the collapse (core-fed; \citealt{McKee2003}) or continue to accrete from the surrounding clump (clump-fed; \citealt{Bonnell2004, VazquezSemadeni2019, Padoan2020}) is still debated, although recent ALMA observations favor a dynamical clump-fed scenario \citep{Motte2018, Sanhueza2019, Coletta2025}.}
 
In the last decade, the advent of interferometers such as the Atacama Large Millimeter Array (ALMA) and the Northern Extended Millimeter Array (NOEMA) resulted in numerous studies of molecular cloud structures on sub-parsec scales. In particular, continuum and molecular-line surveys with ALMA, such as ALMA-IMF \citep{Motte2022} and ALMAGAL \citep{Molinari2025, Sanchez-Monge2025}, have identified pre- and protostellar sources across large samples of high-mass star-forming clumps throughout the Galaxy. Key findings from these programs include an evolution of the core mass function (CMF) \citep{Pouteau2023, Louvet2024}, likely from a Salpeter-like to a top-heavy shape as protoclusters become more dynamically active \citep{Nony2023, Coletta2025}, and a hierarchical, multiscale fragmentation pattern that persists across essentially all evolutionary stages \citep{Molinari2025, Motte2026}. {Furthermore}, clump morphology and fragmentation vary with evolutionary stage and surface density.
Complementary to these statistical, ALMA-based studies, a multi-wavelength investigation of individual high-mass star-forming regions remains necessary. {We therefore selected RCW~108, a massive southern star-forming region that has not yet been extensively studied, to establish whether compact dense-gas structures are associated with its central massive stellar population, and whether their physical properties are consistent with gravitational collapse.}
 
RCW~108 is an embedded \HII\ region within the molecular cloud complex of the Ara~OB1 association, located close to the young open cluster NGC~6193, whose age is estimated to be about 1.0~Myr \citep{Straw1987, Pfalzner2013, Wolk2008}. The Ara~OB1 region shows ongoing star formation \citep{Yamaguchi1999} and is located at a distance of 1.3~kpc \citep{Herbst1977, Kaltcheva1992}.
Previous multiwavelength studies of RCW~108 reported indications of feedback-influenced star formation and a top-heavy initial mass function (IMF) \citep{Comeron2005, Comeron2007}. From the IMF, a total stellar mass of 370~M$_{\odot}$ was estimated. The central region hosts IRS~29, a dust-embedded O9-type star \citep{Urquhart2004}{, which is bright in the X-ray regime ($\rm L_{X}\,\approx\,10^{32.6}\,erg\,s^{-1}$) and associated with water-maser emission \citep{Melnick2000, Smith2002, Wolk2008}. In addition, \cite{Wolk2008} identified about a dozen bright infrared and X-ray sources near IRS~29.}
 
{On scales of about 1-10 pc, RCW~108 is projected close to the filament NGC~6188, which is associated with a rim nebula \citep{Urquhart2004}}. \cite{Comeron2005} presented a detailed analysis of the region using CO millimeter-line observations obtained with the Swedish-ESO Submillimeter Telescope (SEST). However, the limited angular resolution of those data (45$''$ for $^{13}$CO 1-0 and 23$''$ for $^{12}$CO 2-1) does not allow a detailed investigation of the molecular structure of RCW~108, particularly the relationship between the central massive stellar population and the dense gas.
 
In this work, we present an ALMA Band~6 study of compact dense-gas structures in the region of the embedded cluster RCW~108. {We use the continuum emission at 233~GHz to identify compact cores, and the molecular lines C$^{18}$O, $^{13}$CO, H$_2$CO, and CH$_3$OH, extracted at the continuum positions, to investigate their kinematics and physical properties. We address the following questions:} Are some of the continuum-selected cores gravitationally unstable under the adopted assumptions? How are these cores distributed with respect to IRS~29, the central O9-type feedback source? Finally, are their locations and properties consistent with feedback-associated compression of dense gas?
 
The paper is structured as follows. In Sec.~\ref{sec:data_methods}, we describe the observations{, and in Sec.~\ref{sec:methods}, the} analysis methods. In Sec.~\ref{sec:results}, we present the ALMA continuum and molecular-line results, the compact core sample, and their relation to the Chandra/X-ray view of the central stellar population. We discuss the physical interpretation and limitations of the analysis in Sec.~\ref{sec:discussion}, and summarize our conclusions in Sec.~\ref{sec:conclusions}.
\section{Data}
\label{sec:data_methods}

\subsection{ALMA} \label{subsec:alma}

For the millimeter observations presented in this work, we use ALMA band 6 {toward the RCW~108 cluster at R.A.(2000)=16$^h$40$^m$00.10$^s$ and Dec.(2000)=$-$48$^\circ$51$'$45$''$, with a systemic local standard of rest (LSR) velocity of v$_{\rm LSR}=-24.5$~km\,s$^{-1}$, at} frequencies between 218.00 and 233.00 GHz\footnote{The observations (PI: R. Grellmann, Prog. ID: 2016.1.01182.S) can be found on the ALMA science portal \url{https://almascience.eso.org/} as science-ready data.}. The observations cover the emission lines $\rm H_2CO(3_{03}-2_{02})$, {C$^{18}$O\,(2-1),} $\rm ^{13}CO(2-1)$, and $\rm CH_3OH(4-3)$ (Table \ref{tab:h2co_ch2oh_lines}). These molecular lines are observed in a velocity range of about $\pm 20$ km/s with respect to the systemic velocity.  Furthermore, we observed the 233.00 GHz continuum ($\lambda$\,$\sim$1.3\,mm) to assess the presence of dense and compact dust cores.
\begin{table}[t]
	\centering
	\scriptsize
	\setlength{\tabcolsep}{3pt}
	\begin{tabular}{|ccccc|}
		\hline \hline
		   Species & $\nu_{\rm rest}$ & Beam & PA & RMS \\
		           & [GHz] & [arcsec$\times$arcsec] & [$^\circ$] & [mJy/beam] \\ \hline
              ${\rm H_{2}CO}\,(3_{03}-2_{02})$  &  218.22  & 0.96$\times$0.79 & 84.4 & 13.7 \\
              ${\rm H_{2}CO}\,(3_{22}-2_{21})$  &  218.47  & 0.96$\times$0.80 & 84.3 & 13.5 \\
              ${\rm H_{2}CO}\,(3_{21}-2_{20})$  &  218.76  & 0.96$\times$0.79 & 84.0 & 13.3 \\
              CH$_3$OH$\,(4-3)$ & 218.44                   & 0.96$\times$0.79 & 84.5 & 13.6 \\
              C$^{18}$O\,(2-1) & 219.56 & 0.94$\times$0.79 & 91.3 & 8.6 \\
              ${\rm ^{13}CO}\,(2-1)$ &  220.39 & 0.93$\times$0.79 & 91.5 & 9.3 \\ \hline
              Cont. (re-imaged, robust=0.5)  & 233.00 & 0.80$\times$0.66 & 87.4 & 0.39 \\ \hline
              Cont. (archive, robust=0.5) & 233.00 & 0.72$\times$0.63 & 91.9 & 0.34 \\ \hline

	\end{tabular}
	\caption{Observational parameters of line and continuum observations. The rest frequencies, {the synthesized beams,} and the root mean square (RMS) of the noise are indicated. {The C$^{18}$O, H$_2$CO, CH$_3$OH, and $^{13}$CO values refer to the joint 12-m and 7-m cubes (Sec.~\ref{subsec:alma}).} {The line RMS is measured in line-free channels of the primary-beam-corrected cubes where the primary-beam response is $\geq0.9$. For the continuum, the RMS is estimated from the primary-beam-corrected map used for the core extraction (Sec.~\ref{sec:data_core_detection}).}}
	\label{tab:h2co_ch2oh_lines}
\end{table}
The final imaged area of the observations covers a field-of-view (FOV) of 145" $\times$ 145" with a pixel scale of {0.13}" and a restoring beam of {0.72}" (major axis) and {0.63}" (minor axis). We used the 12\,m configuration with a maximum recovered scale of 6.28". {The continuum map and the core extraction are based on these 12-m data. For the C$^{18}$O, H$_2$CO, CH$_3$OH, and $^{13}$CO spectra, we re-imaged the calibrated 12-m and 7-m (ACA) visibilities of the same program jointly, which recovers emission on larger angular scales than the 12-m data alone. Following the Common Astronomy Software Applications for Radio Astronomy \citep[CASA,][]{Casa2022} online guides, the joint cubes were deconvolved as a mosaic with multiscale \texttt{tclean} (robust 0.5) down to four times the thermal noise within a common line mask, on the same spatial grid as the continuum map. Total-power observations are not available, and we discuss this caveat in Sec. \ref{sec:limitation}. The CH$_3$OH and H$_2$CO cubes have a spectral resolution of 0.67~km\,s$^{-1}$, whereas the $^{13}$CO and C$^{18}$O cubes have a corresponding spectral resolution of 1.33~km\,s$^{-1}$.} 
\begin{table}[h!]
	\centering
	\setlength{\tabcolsep}{3pt}
	\begin{tabular}{|cccc|}
		\hline \hline
		   Data type &  Beam & PA & RMS \\
	                &  [arcsec$\times$arcsec] & [$^\circ$] & 10$^{-1}\times$[mJy/beam] \\ \hline
              12m (archive)       & 0.72$\times$0.63  & 91.9 & 3.41  \\
              12m (re-imaged)       & 0.80$\times$0.66  & 87.4 & 3.91 \\
              12m+7m (re-imaged)    & 0.83$\times$0.68  & 87.6 & 7.62  \\ \hline

	\end{tabular}
	\caption{Comparison between the three data sets available for the analysis. We list the related synthesized beam-size of the 12m (archive and re-imaged) and 12m+7m (re-imaged) data. Unless stated otherwise, we use the 12m (archive) and 12m+7m (re-imaged) data for the analysis.}
	\label{tab:comp_data}
\end{table}
{For the analysis, we compare three data sets of the same region (Table~\ref{tab:comp_data}): the public 12m archive data (Fig. \ref{fig:ALMA_fc_12m_archive}), the 12m data re-imaged with CASA, and the combined 12m+7m data. The re-imaged 12m and 12m+7m continuum maps have larger beams and a noise level that is about 15\% and a factor of two higher, respectively, than the 12m archival data. Furthermore, the 12m-archive data contains about twice as many strongly negative pixels (below $-5\,\sigma_{\rm cont}$) as the re-imaged 12m map. 
\begin{figure}[h]
	\centering
	\includegraphics[width=.5\textwidth]{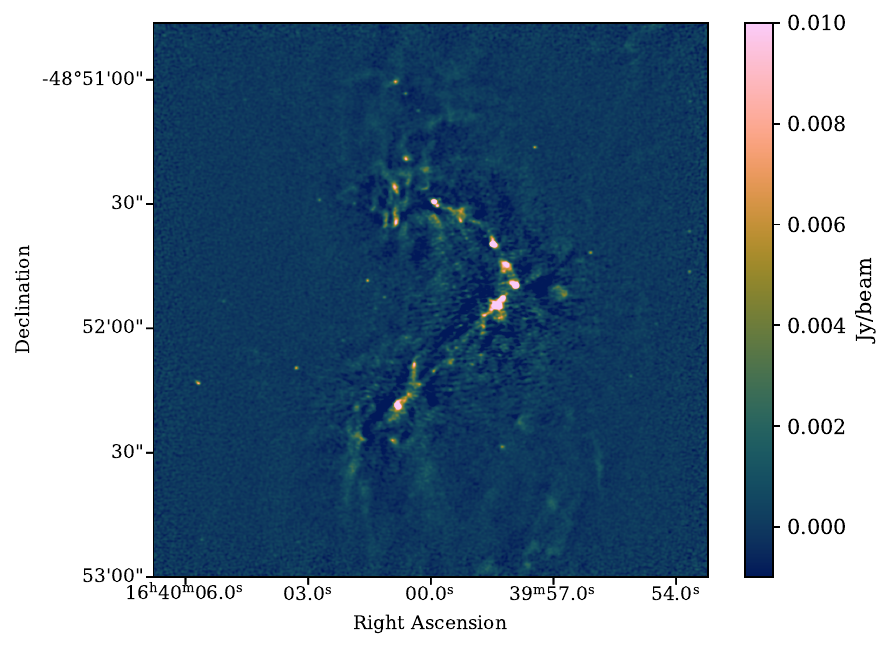}
	\caption{Continuum observation of RCW~108 at 233.00 GHz based on ALMA-archive science ready data. The primary-beam corrected image with a Briggs parameter of robust=0.5 shows negative bowls along the arc-like filaments {(Sec.~\ref{sec:limitation})}.}
\label{fig:ALMA_fc_12m_archive}
\end{figure}
We use the 12m archive map for the continuum core analysis because of its highest angular resolution and lowest noise, and the combined 12m+7m cubes for the kinematic analysis (virial parameters, H$_2$CO temperatures). A similar approach is used in the literature \citep{Sanhueza2019, Ginsburg2022, Pouteau2022, Morii2023, Armante2024, Louvet2024}. We discuss the impact of this choice in Sec.~\ref{sec:limitation}.}

\subsection{Chandra and Spitzer ancillary data} \label{subsec:chandra}

To analyze the location of the massive stars in RCW~108 in relation to the structures observed with ALMA, we {use the Chandra observations obtained with the Advanced CCD Imaging Spectrometer (ACIS-I) and analyzed by} \cite{Wolk2008}. We use science-ready data\footnote{See obs. id 4503 at \url{https://cxc.cfa.harvard.edu/cda/}.} with an exposure time of $\approx$89 ks {in the} energy range of {0.5-7 keV}, and extract an area centered on RCW~108 comparable to the ALMA FOV of 145" $\times$ 145". {Given the hydrogen column density N$_{\rm H}$ estimated for RCW~108 of N$_{\rm H}\,=10^{21}-10^{23}$cm$^{-2}$ \citep{Wolk2008}, the energy range of the Chandra observations ensures detection of embedded sources.}

\begin{figure}[t]
	\centering
	\includegraphics[width=.5\textwidth]{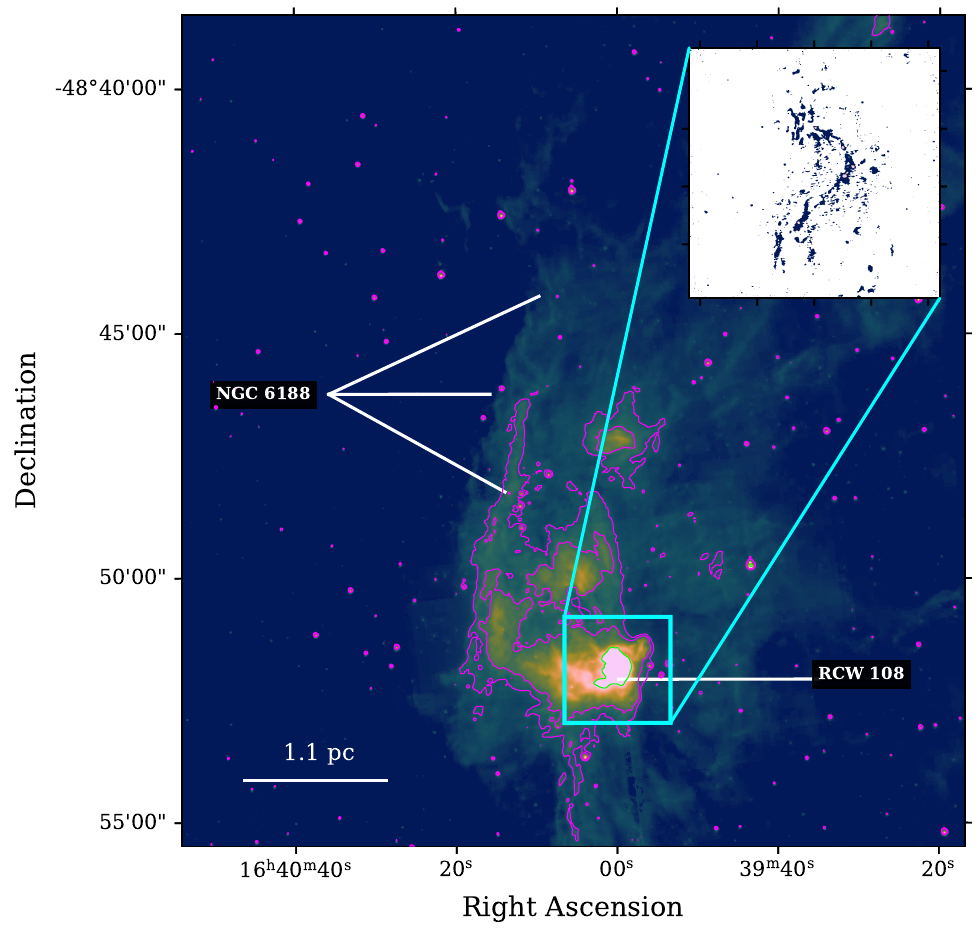}
	\caption{Spitzer mid-infrared finding chart of the RCW~108 region at 8 $\mu$m. See Fig. 1 in \cite{Comeron2007} for a resolved JHK image of the central cluster. In our figure, the rim nebula NGC 6188 extends from North to South. The cluster RCW~108 is located in projection behind NGC 6188. The cyan colored box represents the ALMA FOV and the inset shows the continuum ALMA observations at 233.00 GHz.
    The two pink contour lines show the 1.5\% and 2.5\% level of the 0.23\,Jy peak emission and  the lime colored line shows the 30\% level.
    {The pixel scale is 1.2$^{\prime\prime}$$\times$1.2$^{\prime\prime}$ and the field of view 16$^\prime$$\times$17$^\prime$. North is up, East to the left.}
    }
\label{fig:spitzer_fc}
\end{figure}

Fig.~\ref{fig:spitzer_fc} shows a Spitzer/IRAC 8~$\mu$m finding chart of the region (ProgID 112, PI: G.~Fazio), kindly provided by F.~Comer\'on (priv. comm.) and first presented in \cite{Comeron2007}.

\section{Methodology}
\label{sec:methods}

\subsection{Continuum core identification and molecular-line spectra}
\label{sec:data_core_detection}

The core sample used in this work is defined from the ALMA 233~GHz/1.3 mm continuum emission. We use the PYTHON package \texttt{astrodendro} \citep{Rosolowsky2008} to identify compact continuum structures. {In order to avoid spurious sub-beam detections, we define continuum cores as independent dendrogram leaves (i.e., the smallest structures without further substructure), computed with a minimum intensity of $3\,\sigma_{\rm cont}$, a minimum contrast of $9\,\sigma_{\rm cont}$, and an area comparable to the synthesized beam area of 0.51~arcsec$^2$. Here, $\sigma_{\rm cont}=0.34\,{\rm mJy\,beam^{-1}}$ is the RMS of the primary-beam-corrected continuum map (cf. Table \ref{tab:comp_data}) below the 95th percentile, corresponding to a detection threshold of $3\sigma_{\rm cont}\,=\,1.02\,{\rm mJy\,beam^{-1}}$. All cores used in the CMF and virial analysis lie where the primary-beam response is above 98$\%$ of its peak value, so that the noise of the corrected map at their positions is at most 2$\%$ higher than at the image center.}

The integrated continuum flux density observed at 1.3\,mm is measured from the corresponding \texttt{astrodendro} leaf mask {and converted to Jy with the synthesized beam of the image header}. {A comparison between the primary-beam-corrected and uncorrected map shows that the RMS used for the continuum core analysis is about 1.7 times higher than the thermal noise of 0.21~mJy\,beam$^{-1}$ due to residual extended emission and negative bowls.}

{The size of each core, $r_{\rm obs}=\sqrt{\sigma_{\rm maj}\,\sigma_{\rm min}}$, is given by the intensity-weighted rms size (second moment) of its leaf, as computed by \texttt{astrodendro}. It is deconvolved following $r_{\rm dec}=\sqrt{r_{\rm obs}^2-r_{\rm beam}^2}$ with $r_{\rm beam}=\sqrt{\theta_{\rm maj}\theta_{\rm min}}/(2\sqrt{2\ln2})=0.285''$, and we adopt $R=1.91\,r_{\rm dec}$ as the core radius \citep{Solomon1987}. Cores with $r_{\rm obs}\leq r_{\rm beam}$ are treated as unresolved, with $r_{\rm obs}$ as an upper limit to $r_{\rm dec}$. In total, we identify 33 continuum cores (Table~\ref{tab:continuum_cores}, Appendix~\ref{app:13CO_catalog}).}

{The kinematic information for the virial analysis is derived from the C$^{18}$O\,(2-1) and $^{13}$CO\,(2-1) spectra, extracted directly toward each continuum core and averaged over the pixels of its dendrogram leaf mask. The line emission therefore refers to exactly the same region as the continuum mass. We use C$^{18}$O as the primary tracer, since it is less optically thick than $^{13}$CO and traces denser gas, and $^{13}$CO as an independent tracer to verify the C$^{18}$O results. Each spectrum is fitted with one or two Gaussian components (Appendix~\ref{app:conti_13co_pairs}).}

{The velocity dispersion is corrected for the spectral resolution, $\sigma_{\rm res}=0.566$~km\,s$^{-1}$, for the thermal contribution of the observed molecule, and for the thermal contribution of the mean particle, using
\begin{equation}
\rm \sigma_v^2=\sigma_{\rm obs}^2-\sigma_{\rm res}^2-\frac{k_{\rm B}T}{m_{\rm mol}}+\frac{k_{\rm B}T}{\mu m_{\rm H}},
\label{eq:sigma_v}
\end{equation}
with $T=20$~K and $\mu=2.37$ ($c_s=0.264$~km\,s$^{-1}$). Lines with $\sigma_{\rm obs}<1.2\,\sigma_{\rm res}$ are resolution-limited (RL). For these lines, we use $\max(\sigma_{\rm obs},\sigma_{\rm res})$ without subtracting $\sigma_{\rm res}$, so that $\sigma_v$, $M_{\rm vir}$, and $\alpha_{\rm vir}$ are conservative upper limits. The treatment of the uncertainties is described in Appendix~\ref{sec:uncertainties}, details of the line fitting in Appendix~\ref{app:conti_13co_pairs}.}

\subsection{Core masses and core mass function}

{The masses of the continuum cores are estimated from the integrated 233~GHz flux assuming optically thin dust emission,
}
\begin{equation}
\label{eq:m_i_cont}
       M_{{\rm cont},i} =
       \frac{S_{\nu,i}\,d^2}
       {\kappa_{\nu}\,B_{\nu}(T_{\rm dust})},
\end{equation}
{where $S_{\nu,i}$ is the integrated flux density of the individual continuum core, $d=1.3$~kpc is the distance to RCW~108, $\kappa_{\nu}$ is the gas-dust opacity, and $B_{\nu}(T_{\rm dust})$ is the Planck function. We assume that the 233.00 GHz continuum emission of the cores is dominated by the thermal dust emission. A contribution of free-free emission from ionized gas cannot be excluded for individual cores, but is expected to be negligible for cores embedded in the arc-like filaments. From \cite{Ossenkopf1994}, we adopt $\kappa_{233\,{\rm GHz}}=0.009\,{\rm cm^2\,g^{-1}}$ and use $T_{\rm dust}=20$~K as the reference dust temperature, in agreement with the cold dust traced by SEST \citep{Wolk2008}. Since the high-resolution ALMA data may cover a broader range, we draw $T_{\rm dust}$ between 15 and 30~K in the Monte Carlo analysis (Appendix~\ref{sec:uncertainties}), which changes the masses by factors of 1.48 and 0.60 with respect to 20~K. The assumption of optically thin emission is tested in Appendix~\ref{sec:uncertainties}.}

{To characterize the mass distribution of the continuum cores, we construct the CMF. Although the formal one-beam \(3\sigma_{\rm cont}\) mass sensitivity is approximately \({0.037}\,M_\odot\) for \(T_{\rm dust}=20\)~K, we adopt a conservative lower-mass limit of \({1.0}\,M_\odot\) for the CMF fit, which results in {14} cores observed at 1.3 mm. This threshold is about {27} times higher than the formal one-beam sensitivity and is used to reduce the impact of marginal low-mass detections, source-extraction incompleteness, and spatial filtering.} {We determine the differential slope ${\rm d}N/{\rm d}M\propto M^{-\alpha_{\rm CMF}}$ with the maximum-likelihood estimator $\alpha_{\rm CMF}=1+n\left[\sum_i\ln(M_i/M_{\rm lim})\right]^{-1}$ \citep{Clauset2009}, which does not depend on binning or on the correlated points of a cumulative distribution.} 

{For comparison, we estimate the slope from a linear fit to the cumulative distribution. Based on 5000 iterations, we derive the related uncertainties, each combining one Monte Carlo realization of all core masses (Appendix \ref{sec:uncertainties}) with a bootstrap resampling of the cores above the mass limit.}

\subsection{Virial analysis}

{For each core with index $i$, we derive the virial mass from the velocity dispersion of the C$^{18}$O\,(2-1) line,}
\begin{equation}
    \rm M_{vir, i}\,=\,5\frac{\sigma^2_i R_{i}}{G}
    \label{qe:mass_vir}
\end{equation}
where {$R_i={1.91}\,r_{{\rm dec},i}$ is the beam-corrected radius (Sec.~\ref{sec:data_core_detection}), $\sigma_i$ the velocity dispersion from Eq.~\ref{eq:sigma_v},} and G the gravitational constant with $\rm 4.3\,\times\,10^{-3}\,pc\cdot M^{-1}_{\odot}\cdot (km/s)^2$ \citep[][]{Bertoldi1992, Kauffmann2013}.

{The virial parameter \citep{Bertoldi1992} is the ratio of the virial mass to the continuum mass $M_i$ of each core above the CMF mass limit (Table~\ref{tab:continuum_cores_cmf}),}
\begin{equation}
    \rm \alpha_{\rm vir, i}\,=\,\frac{M_{vir, i}}{M_i}
    \label{eq:alpha_vir}
\end{equation}

{This simplified virial diagnostic of $\alpha_{\rm vir}$ does not include magnetic support or surface-pressure terms, which may be dynamically relevant in the feedback-associated environment of RCW 108.
However, it gives useful information if gravity alone can confine cores and how close these are to virial balance. Low values of $\alpha_{\rm vir}$ are known to be associated with star formation \citep{Kauffmann2013}.
In particular, numerical values for $\alpha_{\rm vir}<2$ suggest supercritical cloud fragments and cores that are prone to collapse. As pointed out by \cite{Kauffmann2013}, very low virial parameters may also indicate that additional support terms, such as magnetic fields or external pressure, are relevant.
}

\subsection{Kinetic temperature using formaldehyde}
\label{sec:data_h2co_line}

{To probe the gas temperature of the cores in the arc-like structures FA and FB (Sec.~\ref{sec:results_yso_rcw108}), we use the para-$\rm H_{2}CO$ transitions $3_{03}-2_{02}$, $3_{22}-2_{21}$, and $3_{21}-2_{20}$ (cf. Table~\ref{tab:h2co_ch2oh_lines}), as well as the line-ratio thermometers introduced by \cite{Magnum1993} and further developed by \cite{Tang2017b, Tang2017a},}
\begin{equation}
\rm T_{kin}\,=\,\frac{47.1\, K}{ln\left(0.556\frac{I_{3_{03}-2_{02}}}{I_{3_{22}-2_{21}}}\right)}\quad and\quad
\rm T_{kin}\,=\,\frac{47.1\, K}{ln\left(0.556\frac{I_{3_{03}-2_{02}}}{I_{3_{21}-2_{20}}}\right)}
\label{eq:kinetic_temperature1}
\end{equation}
{Since H$_2$CO may trace a different gas component than C$^{18}$O, this analysis is not tied to the C$^{18}$O velocity centroids or line widths. To quantify the sensitivity to the extraction, we derive the line ratios for 420 extraction settings per core (cf. Appendix~\ref{app:h2co_details}) and quote the median and the 16-84 percentile range of the valid solutions. Above 150~K, the analytic approximation is not reliable according to the previous authors.}

\subsection{Ionizing photon flux}

{The central star of RCW~108, IRS~29, has been classified as an O9-type source \citep{Urquhart2004, Comeron2005}, without an explicit luminosity class. Since Galactic O9-type stars of luminosity classes I and III have higher $\log Q_0$ values, we adopt luminosity class V as a conservative lower limit for the ionizing photon rate. For an O9V star, \cite{Martins2005} provide $\log Q_0=47.9-48.1$, corresponding to $Q_0\sim10^{48}\,{\rm s^{-1}}$.} The incident ionizing photon flux at the position of each core is obtained as
\begin{equation}
    \Phi_i = \frac{Q_0}{4\pi d_{{\rm proj},i}^2},
    \label{eq:photon_flux}
\end{equation}
where \(d_{{\rm proj},i}\) is the projected separation from IRS~29. {Eq.~\ref{eq:photon_flux} does not account for the absorption of the ionizing photons along their path. We discuss this limitation in Sec.~\ref{sec:disc_irs29}.}

\section{Results}
\label{sec:results}

\subsection{Continuum core detection in RCW~108}
\label{sec:Continuum_detection}

To investigate compact cores in RCW~108, we analyze the ALMA continuum emission at 233~GHz and shown in Fig. \ref{fig:ALMA_fc_12m_archive}. 
\begin{figure}[h!]
	\centering
	\includegraphics[width=.5\textwidth]{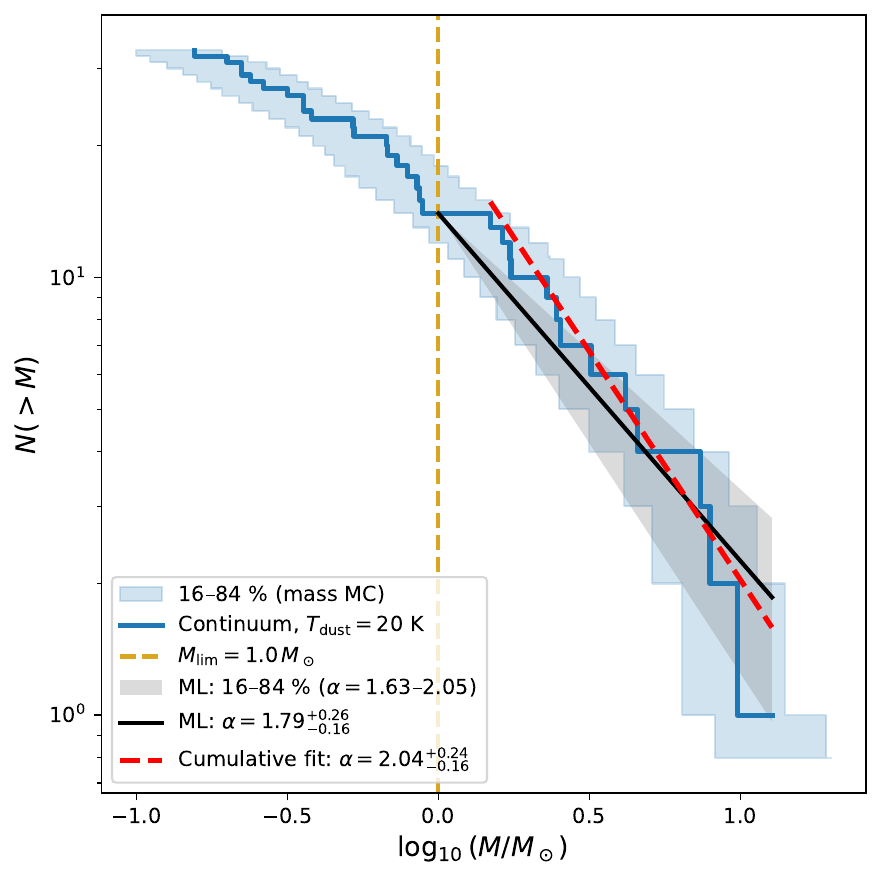}
	\caption{Cumulative core mass function of the continuum-selected cores in RCW~108. In this representation, we show the uncertainty of the masses (blue shaded) and of the fit (grey shaded). With a uniform temperature of $T_{\rm dust}=20$~K for all cores (blue thick line), the red dashed line corresponds to a linear fit to the cumulative distribution with a slope of $\alpha_{\rm CMF}={2.04}$. With $10^4$ Monte Carlo realizations, the blue shaded uncertainty represents core to core temperature variations between 15-30 K, flux and calibration noise, a 2$\sigma$ dust opacity variance, and a gas to dust ratio between 100-110. The black line shows the maximum-likelihood power law above the mass limit ($\alpha_{\rm CMF}={1.79}$) of M$_{\rm lim}\,=\,1.0\,$M$_{\odot}$ (vertical dashed line). Its uncertainty (grey shaded) is derived from 5000 iterations, each combining one Monte Carlo realization with a bootstrap resampling of the cores.}
\label{fig:cmf_continuum}
\end{figure}
\begin{table*}[h!]
\centering
\caption{Continuum cores above the adopted CMF lower-mass limit of $\rm 1\,M_\odot$ based on the 233.00 GHz observations.}
\label{tab:continuum_cores_cmf}
\begin{tabular}{cccccccc}
\hline\hline
ID & R.A. & Dec & Flux & $M_{20\,{\rm K}}$ & $M$ (MC) & $r_{\rm obs}$ & $R$ \\
   & (deg) & (deg) & ($10^{-2}$ Jy) & (M$_\odot$) & (M$_\odot$) & (arcsec) & (arcsec) \\
\hline
22 (B-5) & 249.9934 & -48.8651 & $35.23\pm0.09$ & 12.72 & $11.72^{+6.53}_{-4.11}$ & $0.40\pm0.01$ & 0.53$\pm$0.04 \\
4 & 250.0032 & -48.8718 & $27.14\pm0.24$ & 9.80 & $9.05^{+5.06}_{-3.24}$ & $1.18\pm0.02$ & 2.18$\pm$0.04 \\
25 (B-3) & 249.9916 & -48.8638 & $21.97\pm0.14$ & 7.93 & $7.33^{+4.01}_{-2.61}$ & $0.59\pm0.02$ & 0.99$\pm$0.04 \\
29 (B-1) & 249.9937 & -48.8610 & $20.37\pm0.14$ & 7.35 & $6.66^{+4.01}_{-2.30}$ & $0.53\pm0.04$ & 0.85$\pm$0.09 \\
27 (B-2) & 249.9924 & -48.8626 & $12.64\pm0.17$ & 4.57 & $4.20^{+2.44}_{-1.49}$ & $0.88\pm0.01$ & 1.60$\pm$0.03 \\
36 (A-1) & 249.9997 & -48.8582 & $11.52\pm0.10$ & 4.16 & $3.85^{+2.11}_{-1.37}$ & $0.44\pm0.02$ & 0.63$\pm$0.04 \\
35 (A-2) & 249.9973 & -48.8590 & $8.86\pm0.17$ & 3.20 & $2.94^{+1.66}_{-1.03}$ & $1.10\pm0.02$ & 2.03$\pm$0.03 \\
37 & 250.0040 & -48.8571 & $7.01\pm0.19$ & 2.53 & $2.34^{+1.31}_{-0.83}$ & $1.27\pm0.52$ & 2.36$\pm$1.02 \\
30 & 249.9995 & -48.8598 & $6.81\pm0.19$ & 2.46 & $2.27^{+1.25}_{-0.81}$ & $1.30\pm0.34$ & 2.42$\pm$0.67 \\
33 & 250.0036 & -48.8593 & $6.33\pm0.14$ & 2.29 & $2.09^{+1.22}_{-0.73}$ & $0.74\pm0.02$ & 1.31$\pm$0.04 \\
0 & 250.0076 & -48.8740 & $4.82\pm0.17$ & 1.74 & $1.60^{+0.87}_{-0.57}$ & $1.02\pm0.51$ & 1.88$\pm$1.01 \\
24 & 249.9868 & -48.8643 & $4.78\pm0.16$ & 1.73 & $1.59^{+0.90}_{-0.56}$ & $0.90\pm0.04$ & 1.63$\pm$0.09 \\
23 (B-4) & 249.9928 & -48.8647 & $4.51\pm0.05$ & 1.63 & $1.50^{+0.82}_{-0.53}$ & $0.28\pm0.01$ & $\lesssim$0.54 \\
8 & 250.0018 & -48.8694 & $4.12\pm0.10$ & 1.49 & $1.37^{+0.77}_{-0.49}$ & $0.59\pm0.26$ & 0.98$\pm$0.56 \\
\hline
\end{tabular}
\tablefoot{Fluxes from the dendrogram leaf masks with the header beam and a statistical flux error given as $\sigma_{\rm cont}/{\rm PB}_i\sqrt{N_{\rm pix}/N_{\rm beam}}$. For the core mass, we assume a dust temperature of $T_{\rm dust}=20$~K and a gas-dust opacity of $\kappa_{233\,{\rm GHz}}=0.009\,{\rm cm^2\,g^{-1}}$. {$M$ (MC) gives the Monte Carlo median and the 16-84 percentile range uncertainty (cf. Appendix~\ref{sec:uncertainties}).} The intensity-weighted second-moment radius of the leaf {is} reported as $r_{\rm obs}$. The radius R is calculated from the beam radius r$_{\rm beam}$ and the observed radius $r_{\rm obs}$ (cf Sec. \ref{sec:methods} for details).}
\end{table*}
{Using the method described in Sec.~\ref{sec:data_core_detection}}, we identify {33} continuum cores. The detected sources have integrated fluxes {ranging} between ${4.32}\times10^{-3}$ and ${0.352}$~Jy and {observed sizes {$r_{\rm obs}$} between ${0.22}''$ and ${1.30}''$}.

Table \ref{tab:continuum_cores_cmf} lists the 14 continuum cores above the lower-mass limit of ${1.0}\,M_\odot$\footnote{The complete catalog is given in Table~\ref{tab:continuum_cores} (Appendix~\ref{app:13CO_catalog}).}. For the reference temperature of $T_{\rm dust}=20$~K, the continuum core masses range from ${0.16}$ to ${12.72}\,M_\odot$, with a median mass of ${0.85}\,M_\odot$.

{Among these 14 cores, we find four (4, 22, 25, 29) with masses of $\approx$7.4-12.7~M$_\odot$. For the brightest core 22, the optical depth reaches $\tau\approx1.1$, so that its mass may be underestimated by up to a factor of $\approx$1.6. For all other cores, we find $\tau\leq0.5$ (Appendix~\ref{sec:uncertainties}).}

The total mass contained in the {33 continuum cores is $M_{\rm total}=69.4^{+29.9}_{-21.3}\,M_\odot$, in agreement with the estimates for the RCW~108 core region in \cite{Comeron2005}. The cumulative CMF is shown in Fig.~\ref{fig:cmf_continuum}. For the differential power law ${\rm d}N/{\rm d}M \propto M^{-\alpha_{\rm CMF}}$ above $1.0\,M_\odot$, the maximum-likelihood (ML) estimate is $\alpha_{\rm CMF}=1.79^{+0.26}_{-0.16}$ whereas the linear fit to the cumulative distribution results in $\alpha_{\rm CMF}=2.04^{+0.24}_{-0.16}$. Both estimates agree within their uncertainties. A steeper slope of the cumulative fit is expected, since the points of a cumulative distribution are not independent and the fit is dominated by the few most massive cores. We finally adopt the maximum-likelihood estimate $\alpha_{\rm CMF}=1.79^{+0.26}_{-0.16}$ where the uncertainty is derived from the Monte Carlo and bootstrap analysis.} 

{In the following, all values are given as Monte Carlo medians with 16-84 percentile ranges (Appendix~\ref{sec:uncertainties}), except for the reference masses $M_{20\,{\rm K}}$ at $T_{\rm dust}=20$~K.}

\subsection{Two regions of interest in RCW~108}
\label{sec:regionA}
\begin{figure}[h!]
	\centering
	\includegraphics[width=.5\textwidth]{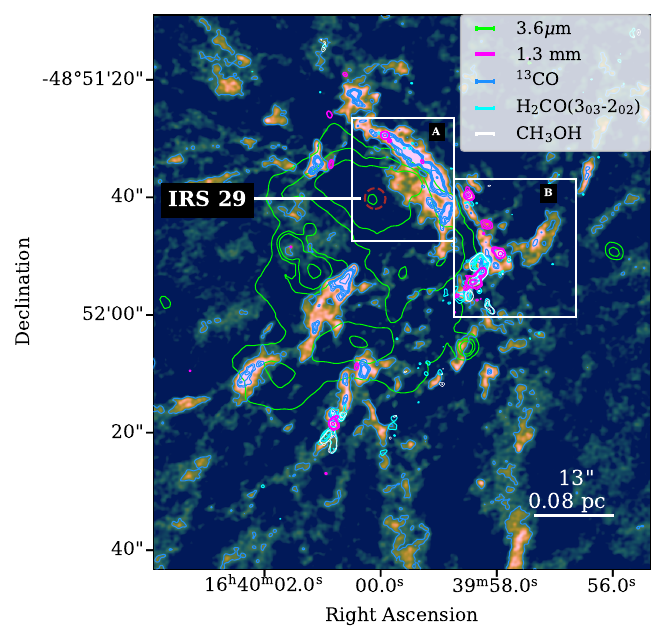}
	\caption{{Composite finding chart of RCW~108 based on ALMA 12m archive data and Spitzer observations. The image shows the ${\rm H_{2}CO}\,(3-2)$, ${\rm ^{13}CO}\,(2-1)$, and CH$_3$OH$\,(4-3)$ emission. IRS~29 is marked with a dashed brown circle. Contour levels are given in percent of the respective peak: Spitzer 3.6~$\mu$m at 5, 10, 16, and 75\% of 47~mJy/arcsec$^2$ (IRAC data of the same program as in Fig.~\ref{fig:spitzer_fc}), 233~GHz continuum (magenta) at 3, 6, 28, 56, and 100\% of 0.19~Jy/beam. Collapsed $^{13}$CO cube at 15, 44, 50, 56, 62, and 94\% of 0.16~Jy/beam~km/s. H$_2$CO at 17, 34, 50, and 67\% of 0.06~Jy/beam~km/s. CH$_3$OH at 14 and 19\% of 0.21~Jy/beam~km/s. The white boxes mark the regions A and B.}}
\label{fig:rcw108_finding_chart}
\end{figure}
Figure~\ref{fig:rcw108_finding_chart} shows a composite ALMA line-emission image of RCW~108.
{It combines the integrated-intensity maps of $^{13}{\rm CO}$, H$_2$CO, and CH$_3$OH with the 233~GHz continuum and the 3.6~$\mu$m emission from Spitzer/IRAC, tracing the embedded stellar population \citep{Comeron2005, Wolk2008}.
With this multi-line approach, we identify two arc-like ${\rm ^{13}CO}$ features marked with `A' and `B' that show embedded compact 1.3 mm continuum emission.} 

\begin{figure*}[htbp!]
	\centering
	\includegraphics[width=.49\textwidth]{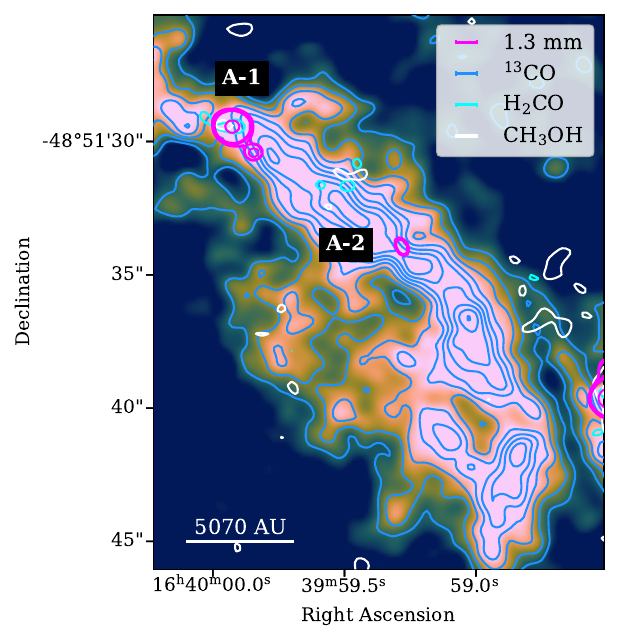}\hfill
	\includegraphics[width=.49\textwidth]{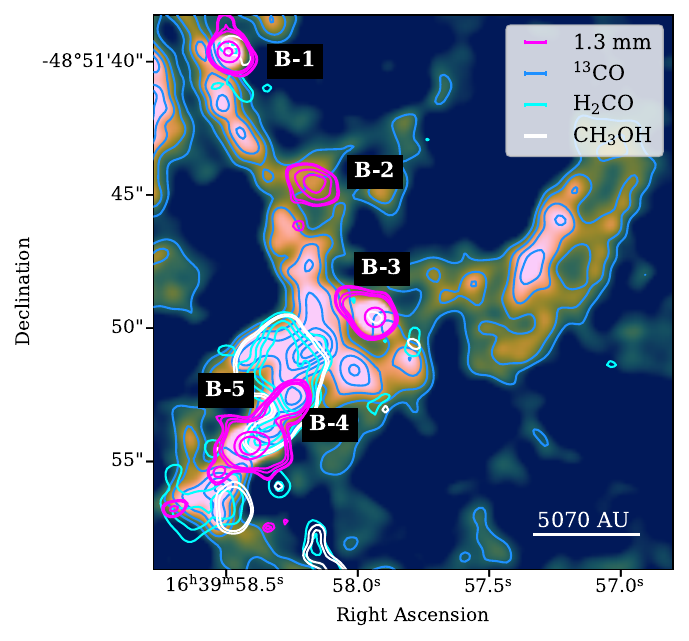}
	\caption{{Left:} Close view of the filament FA partially enveloping the central region of RCW~108 (region A in Fig.~\ref{fig:rcw108_finding_chart}). The two continuum cores A-1 and A-2 are identified. {Right:} Close view of the arc-like filament FB (region B in Fig.~\ref{fig:rcw108_finding_chart}). The continuum cores B-1, B-2, B-3, B-4, and B-5 are identified. In both panels, we show the molecular emission lines $^{13}$CO, H$_2$CO\,(3-2), and CH$_3$OH\,(4-3) with contour lines. For $^{13}$CO, the contour lines are chosen in steps of 10\% of the 0.16~Jy/beam peak emission. For the other lines, the contour levels are as in Fig.~\ref{fig:rcw108_finding_chart}.}
\label{fig:rcw108_boxes}
\end{figure*}

In region A, we observe an arc-like ${\rm ^{13}CO}$ filament (FA) at a projected separation of $\approx$10.5$''$ ($\approx1.4\times10^4$~AU) north-west of IRS~29 (Figs.~\ref{fig:rcw108_finding_chart} and \ref{fig:rcw108_boxes}, left). The continuum emission reveals the presence of the dense cores A-1 (ID-36) and A-2 (ID-35) embedded in FA. For A-1, we also detect ${\rm H_{2}CO}\,(3-2)$ and CH$_3$OH\((4-3)\) {molecular} emission. The latter one may be indicative of shocked gas, suggesting the presence of outflows associated with the source. The comparison of its position with the protostellar sources identified in \cite{Comeron2007} shows that A-1 can be associated with CS-109, a strongly embedded candidate protostellar object. 

{At a larger projected separation of $\approx$16-23$^{\prime\prime}$ from IRS29, region B (Fig.~\ref{fig:rcw108_finding_chart}) contains a second arc-like ${\rm ^{13}CO}$ filament (FB) with an angular size comparable to FA (Fig.~\ref{fig:rcw108_boxes}, right). Along FB, we detect five compact continuum cores identified as B-1, B-2, B-3, B-4 and B-5 with a relative projected separation of $\approx$6$''$ ($\approx$8$\times$10$^3$~AU, Table~\ref{tab:spaced_fragemention}). H$_2$CO emission is detected toward B-1, B-2, B-3, and B-5, and B-1 shows compact H$_2$CO and CH$_3$OH emission at the continuum position (Sec. \ref{sec:H2CO_temp}).}

\begin{table}[h!]
	\centering
	\begin{tabular}{|cccc|}
		\hline \hline
	   Cores         & \multicolumn{3}{c|}{Neighbor Distance}  \\
  		              & [arcsec]          & [AU]  & [pc] \\ \hline
    B-1/B-2    &    5.9            & 7670  & 0.037\\
    B-2/B-3    &    5.6            & 7280  & 0.035 \\
    B-3/B-5   &    6.5            & 8450  & 0.041 \\ \hline
	\end{tabular}
	\caption{Projected {physical separation} between neighboring cores in {FB}. The average {separation} is 7800 AU (or 0.038 pc). B-4 is omitted because of its small separation from B-5.}
	\label{tab:spaced_fragemention}
\end{table}

\subsection{The {X-ray luminous} massive stars in RCW\,108} 
\label{sec:stars}

\begin{figure}[htbp!]
	\centering
	\includegraphics[width=.5\textwidth]{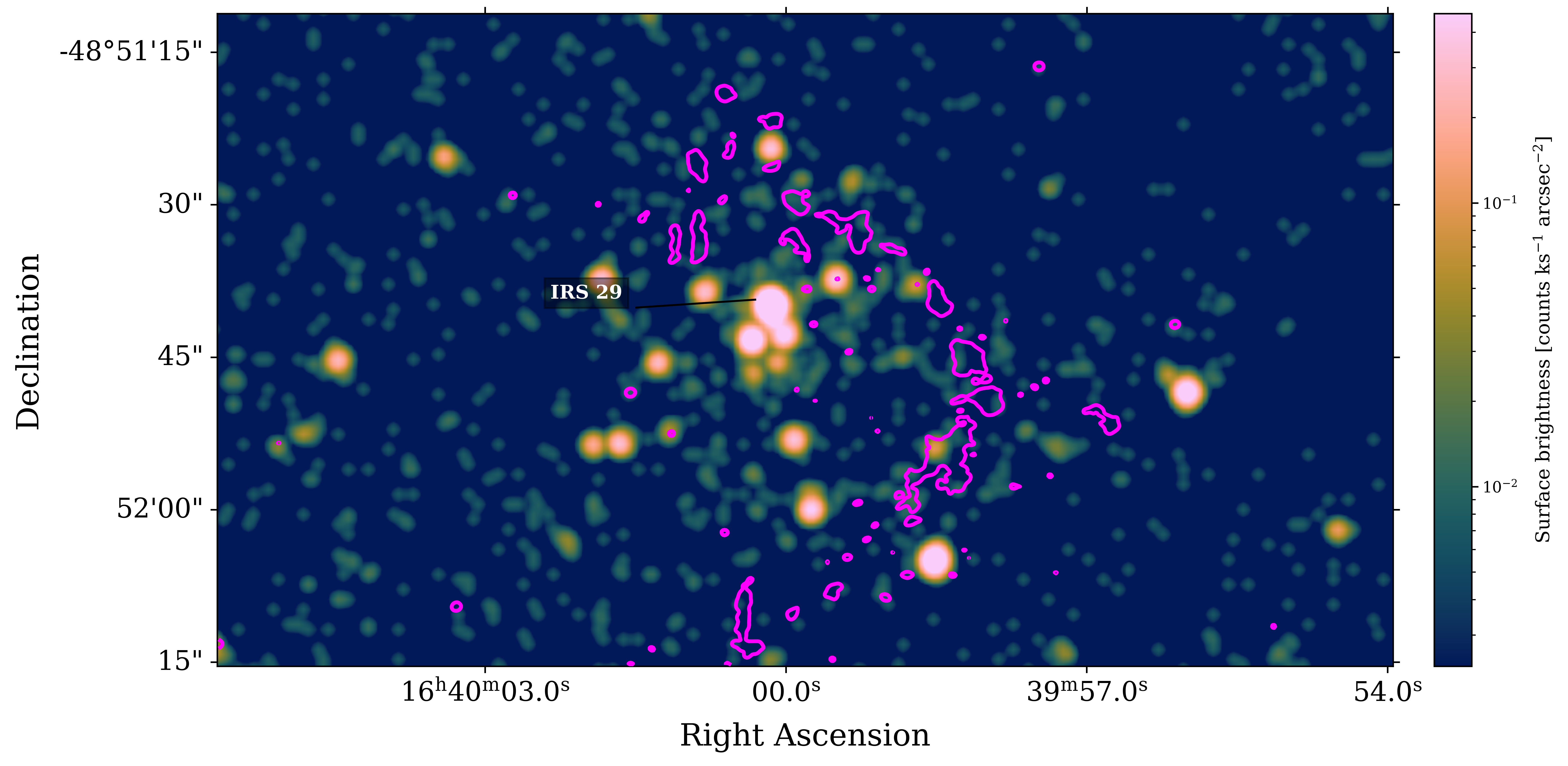}
	\caption{Finding chart of the massive stars in RCW~108 region observed with Chandra (0.5-7 keV). We show the one-pixel Gaussian smoothed X-ray emission overlaid with the continuum flux observed with ALMA at 233.00 GHz. The units of the Chandra data are in surface brightness. The contour level refers to 1$\%$ of the peak flux at 233.00 GHz of {0.19} Jy/beam. The brightest X-ray source in the region is the {massive} O9-type star IRS~29 \citep{Straw1987, Comeron2007}.}
\label{fig:chandra_fc_rcw108}
\end{figure}

We compare the spatial distribution of the compact cores observed at 1.3 mm to the Chandra (0.5-7 keV) X-ray emission tracing the presence of high-mass stars (Fig.~\ref{fig:chandra_fc_rcw108}). A visual inspection shows that the massive O9 star IRS\,29 and several of the massive stars revealed by Chandra are found inside the approximate circular region delimited by the arc-like continuum emission, whereas other stars are found outside this region. A quantitative analysis of the stellar content of RCW\,108 will be covered in a forthcoming study (Nagar et al., in prep.).

\subsection{Virial properties of the continuum cores}
\label{sec:virial_analysis}

{To test whether the 14 cores above 1.0~M$_\odot$ (Table~\ref{tab:continuum_cores_cmf}) are prone to collapse, we use the C$^{18}$O line widths to derive their virial parameters (Eqs.~\ref{qe:mass_vir} and \ref{eq:alpha_vir}).}
\begin{figure}[h!]
	\centering
	\includegraphics[width=0.5\textwidth]{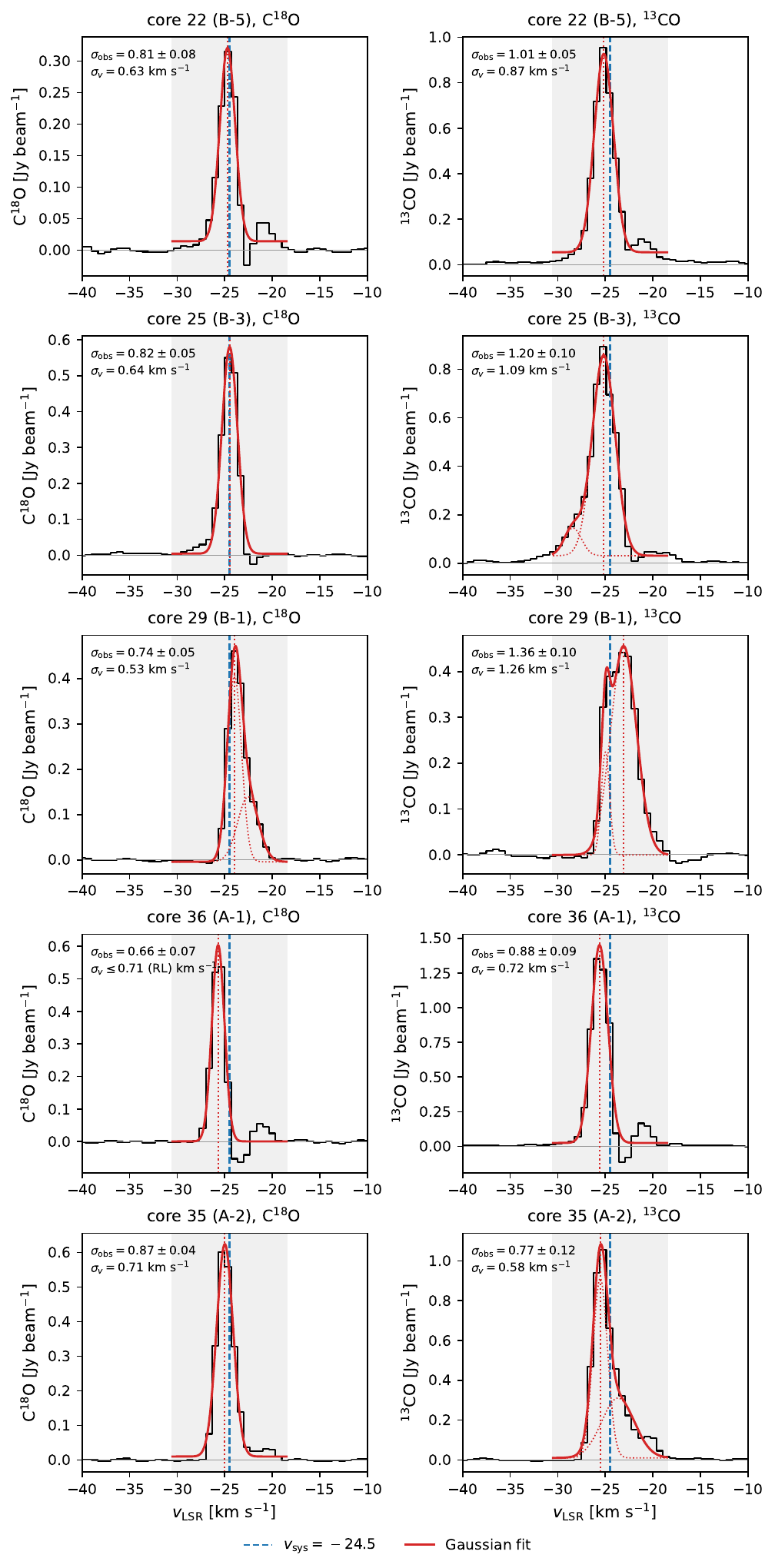}
	\caption{{C$^{18}$O\,(2-1) (left) and $^{13}$CO\,(2-1) (right) spectra of the cores 22 (B-5), 25 (B-3), 29 (B-1), 36 (A-1), and 35 (A-2) in the filaments FA and FB, averaged over the continuum leaf masks. The red lines show the Gaussian fits (dotted: individual components of two-component fits) within the grey fit window $v_{\rm sys}\pm6$~km\,s$^{-1}$ whereas the dotted red vertical line marks the fitted line center. The dashed blue line marks the systemic velocity. The observed width $\sigma_{\rm obs}$ and the derived velocity dispersion $\sigma_v$ are given in each panel.}} 
\label{fig:spectra_main}
\end{figure}

{C$^{18}$O emission is detected for all 14 cores, with peak signal-to-noise (SNR) ratios between 61 and 351 in the spectra averaged over the leaf masks. In addition, we detect $^{13}$CO for all cores as well, with SNR between 74 and 657. Spectra of the cores in FA and FB are shown in Fig.~\ref{fig:spectra_main}, while all spectra are reported in Appendix~\ref{app:conti_13co_pairs}. For two of the 14 cores, the observed C$^{18}$O line width is resolution-limited (RL, Sec.~\ref{sec:data_core_detection}), so that $\sigma_v$, $M_{\rm vir}$, and $\alpha_{\rm vir}$ are upper limits. For $^{13}$CO, this applies to one core. The $^{13}$CO lines are mostly broader than the C$^{18}$O lines, as expected for a more optically thick tracer of extended gas.}

{In the combined 12-m and 7-m cubes, the spectra are largely free of the negative bowls that are produced by interferometric filtering of extended emission in 12-m-only data. For a few cores, narrow negative features remain on the red side of the C$^{18}$O line (e.g., cores 22 and 36, Fig.~\ref{fig:spectra_main}), while the fitted constant baselines are close to zero. We discuss these effects and their impact on the data in Sec.~\ref{sec:limitation}.}

\begin{figure}[h!]
	\centering
	\includegraphics[width=0.5\textwidth]{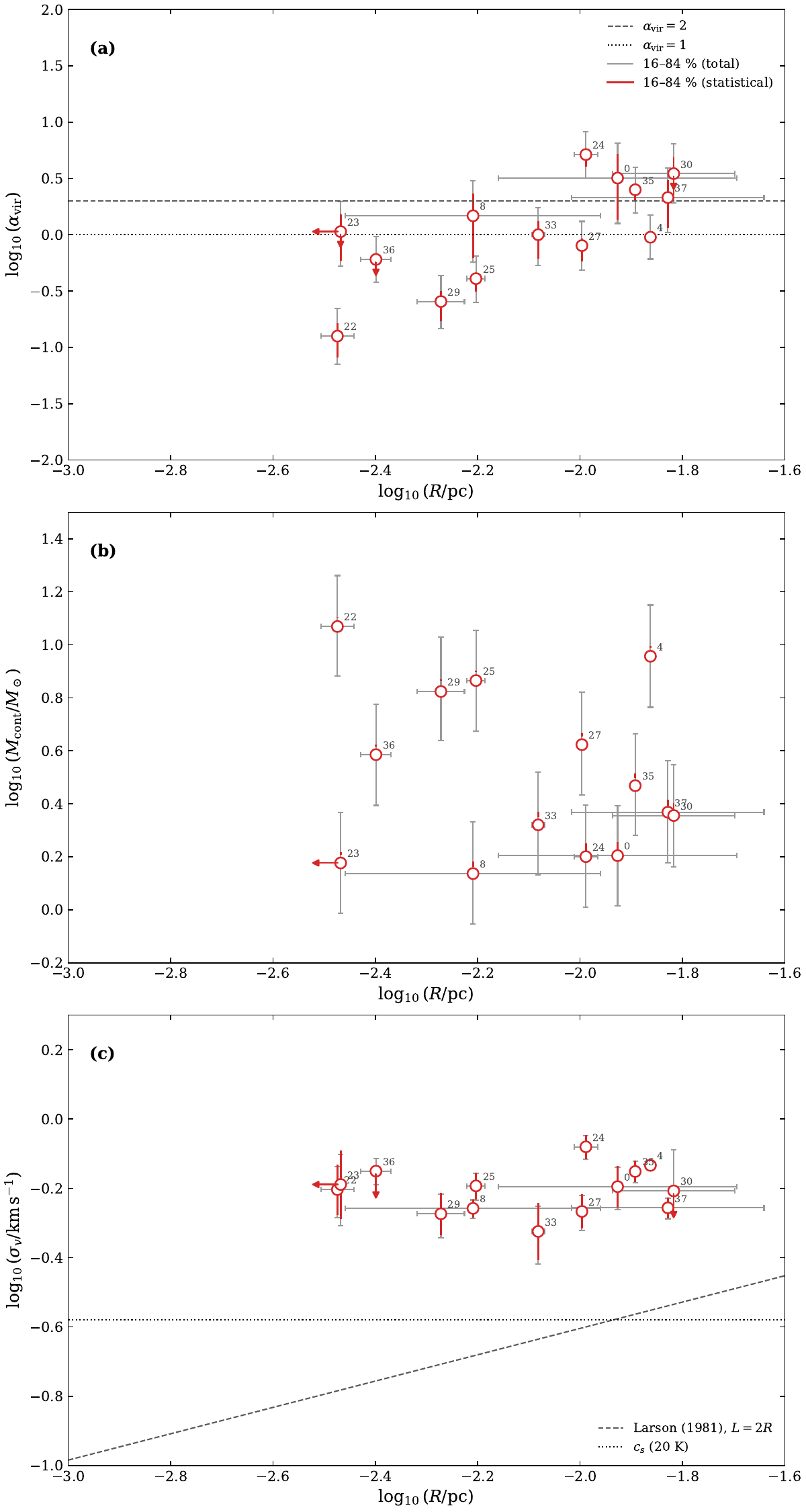}
	\caption{{Virial analysis of the 14 continuum cores above $1.0\,M_\odot$ with C$^{18}$O emission line widths. Panel a) shows the virial parameter $\alpha_{\rm vir}$ as a function of the radius $R=1.91\,r_{\rm dec}$. Below the threshold of $\alpha_{\rm vir}=2$, the cores are consistent with being gravitationally bound or prone to collapse. Panel b) represents the continuum mass as a function of $R$. Panel c) shows the linewidth-size relation, i.e., the C$^{18}$O velocity dispersion $\sigma_v$ as a function of $R$. The dashed line represents the relation of \cite{Larson1981} for a size $L=2R$, the dotted line the sound speed at 20~K. Symbols show the Monte Carlo medians (panel c: fitted $\sigma_v$), grey error bars give the 16-84 percentile range of the total uncertainty, and red error bars the statistical uncertainty only (Table~\ref{tab:uncertainty_budget}). Arrows indicate upper limits. A downward direction represents a resolution-limited line and/or an unresolved radius (panel a) or a resolution-limited line (panel c). The leftward direction indicates unresolved radii.}}
\label{fig:virial_parameter_mass}
\end{figure}

The resulting virial parameter $\rm \alpha_{\rm vir, i}$ of the individual cores as a function of radius is shown in Fig. \ref{fig:virial_parameter_mass} {and listed, together with the uncertainties and $P(\alpha_{\rm vir}<2)$, in Table~\ref{tab:continuum_cores_cmf_virial}. Of the 14 cores, 9 have $\alpha_{\rm vir}<2$. For an RL core, an upper limit below two is a robust indication that the core is bound under the adopted assumptions.}
{The five cores with $\alpha_{\rm vir}\geq2$ (35, 37, 30, 0, 24) have resolved C$^{18}$O lines, with the exception of core 30, whose line is resolution-limited, so that its value is an upper limit. Their dynamical state is discussed in Sec.~\ref{sec:disc_sf}.}

{Figure~\ref{fig:virial_parameter_mass}c shows the linewidth-size relation of the cores. With $\sigma_v\approx0.5-0.8$~km\,s$^{-1}$, i.e., about 2-3 times the sound speed, all cores lie a factor of $\approx$2-4 above the relation of \cite{Larson1981}. Over the covered range of $R\approx0.003-0.015$~pc, the C$^{18}$O line widths do not depend on the core size.}
\begin{table*}[htpb]
\centering
\caption{Kinematic and virial properties of the continuum cores above $1\,M_\odot$ using the C$^{18}$O molecular line.} 
\label{tab:continuum_cores_cmf_virial}
\setlength{\tabcolsep}{4pt}
\begin{tabular}{cccccccccc}
\hline\hline
ID & $M_{20\,{\rm K}}$ & $R$ & $\Delta v_{\rm LSR}$ & $\sigma_{\rm obs}$ & $\sigma_v$ (C$^{18}$O) & $\sigma_v$ ($^{13}$CO) & $M_{\rm vir}$ & $\alpha_{\rm vir}$ & $P(\alpha_{\rm vir}<2)$ \\
 & (M$_\odot$) & (pc) & (km\,s$^{-1}$) & (km\,s$^{-1}$) & (km\,s$^{-1}$) & (km\,s$^{-1}$) & (M$_\odot$) & & \\
\hline
22 (B-5) & 12.72 & 0.0034 & $-0.20$ & $0.81\pm0.08$ & $0.63\pm0.11$ & $0.87\pm0.06$ & $1.53^{+0.54}_{-0.49}$ & $0.13^{+0.09}_{-0.05}$ & 1.00 \\
4 & 9.80 & 0.0137 & $+0.20$ & $0.89\pm0.01$ & $0.74\pm0.02$ & $1.10\pm0.08$ & $8.62^{+0.40}_{-0.38}$ & $0.95^{+0.54}_{-0.34}$ & 0.96 \\
25 (B-3) & 7.93 & 0.0063 & $+0.02$ & $0.82\pm0.05$ & $0.64\pm0.06$ & $1.09\pm0.11$ & $3.00^{+0.56}_{-0.52}$ & $0.41^{+0.24}_{-0.16}$ & 1.00 \\
29 (B-1) & 7.35 & 0.0053 & $+0.53$ & $0.74\pm0.05$ & $0.53\pm0.08$ & $1.26\pm0.11$ & $1.75^{+0.57}_{-0.50}$ & $0.26^{+0.18}_{-0.11}$ & 1.00 \\
27 (B-2) & 4.57 & 0.0101 & $+0.01$ & $0.74\pm0.05$ & $0.54\pm0.06$ & $0.71\pm0.06$ & $3.43^{+0.82}_{-0.77}$ & $0.80^{+0.51}_{-0.32}$ & 0.97 \\
36 (A-1) & 4.16 & 0.0040 & $-1.19$ & $0.66\pm0.07$ & $\lesssim$0.71$^{\rm RL}$ & $0.72\pm0.11$ & $\lesssim2.32^{+0.46}_{-0.39}$ & $\lesssim0.60^{+0.36}_{-0.23}$ & $\geq$0.99 \\
35 (A-2) & 3.20 & 0.0128 & $-0.49$ & $0.87\pm0.04$ & $0.71\pm0.05$ & $0.58\pm0.16$ & $7.43^{+1.08}_{-1.05}$ & $2.51^{+1.44}_{-0.94}$ & 0.32 \\
37 & 2.53 & 0.0148 & $+0.61$ & $0.75\pm0.03$ & $0.55\pm0.04$ & $0.89\pm0.04$ & $5.26^{+2.46}_{-2.32}$ & $2.14^{+1.80}_{-1.09}$ & 0.46 \\
30 & 2.46 & 0.0152 & $-1.15$ & $0.56\pm0.22$ & $\lesssim$0.62$^{\rm RL}$ & $\lesssim$0.72$^{\rm RL}$ & $\lesssim7.77^{+4.21}_{-2.32}$ & $\lesssim3.49^{+2.92}_{-1.58}$ & $\geq$0.18 \\
33 & 2.29 & 0.0083 & $+0.07$ & $0.69\pm0.06$ & $0.47\pm0.09$ & $0.46\pm0.14$ & $2.15^{+0.86}_{-0.76}$ & $1.00^{+0.73}_{-0.47}$ & 0.89 \\
0 & 1.74 & 0.0118 & $-0.33$ & $0.81\pm0.07$ & $0.64\pm0.09$ & $0.90\pm0.16$ & $5.30^{+3.78}_{-2.94}$ & $3.19^{+3.30}_{-1.93}$ & 0.29 \\
24 & 1.73 & 0.0103 & $+1.42$ & $0.97\pm0.06$ & $0.83\pm0.07$ & $0.52\pm0.13$ & $8.23^{+1.45}_{-1.32}$ & $5.15^{+3.05}_{-1.97}$ & 0.02 \\
23 (B-4) & 1.63 & $\lesssim$0.0034 & $-0.41$ & $0.82\pm0.12$ & $0.65\pm0.15$ & $1.06\pm0.09$ & $\lesssim1.66^{+0.81}_{-0.70}$ & $\lesssim1.07^{+0.91}_{-0.54}$ & $\geq$0.85 \\
8 & 1.49 & 0.0062 & $+0.12$ & $0.75\pm0.02$ & $0.55\pm0.03$ & $0.94\pm0.06$ & $2.15^{+1.31}_{-1.24}$ & $1.47^{+1.52}_{-0.90}$ & 0.67 \\
\hline
\end{tabular}
\tablefoot{Spectra averaged over the continuum leaf mask and a corresponding Gaussian fit within $v_{\rm sys}\pm6$~km\,s$^{-1}$. If we identify two components, the brighter one is adopted. The fit errors are based on covariance which are scaled by $\sqrt{2}$ for the Hanning-correlated channels. The resolution-limited cores are indexed with RL. For M$_{\rm vir}$ and $\alpha_{\rm vir}$, the medians and the 16-84 percentile range are based on $10^4$ Monte Carlo realizations. The fraction of runs with $\alpha_{\rm vir}<2$ in the $10^4$ Monte Carlo realizations is given in the last column. Lower ($\geq$) and upper ($\lesssim$) limits are indicated wherever applicable.}
\end{table*}

{Using $^{13}$CO instead of C$^{18}$O (Appendix~\ref{app:conti_13co_pairs}) or $R$=1.177\,$r_{\rm dec}$ (half of the deconvolved FWHM) instead of 1.91$\,r_{\rm dec}$ does not change the qualitative picture for the cores in FA and FB. We find two exceptions, core 23 (B-4) and core 35 (A-2), that are discussed in Sec. \ref{sec:disc_sf}.}

\begin{figure*}[htpb]
	\centering
	\includegraphics[width=1.0\textwidth]{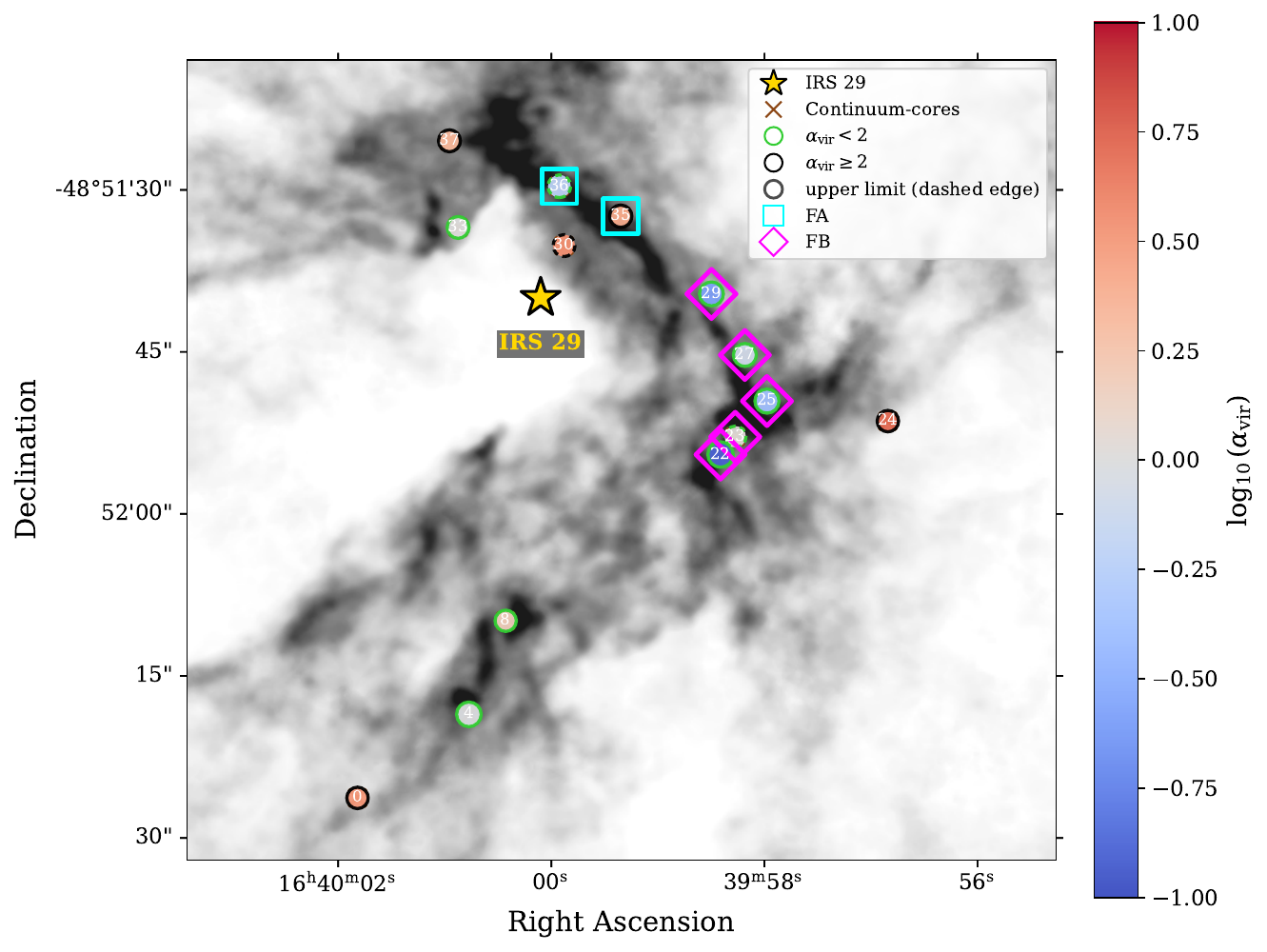}
	\caption{Location of all {14} continuum cores that are listed in Table \ref{tab:continuum_cores_cmf}. The background image shows a collapsed $\rm ^{13}CO$ cube {based on the re-imaged 12m-only data listed in Table \ref{tab:comp_data}}. The continuum cores are marked with a brown cross. {The virial parameter $\alpha_{\rm vir}$ based on the C$^{18}$O molecular line is color-coded (Fig. \ref{fig:virial_parameter_mass}). Cores with $\alpha_{\rm vir}<2$ have a green edge, cores with $\alpha_{\rm vir}\geq2$ have a black edge. Dashed edges mark upper limits.} This representation shows that the majority of cores that are prone to collapse are located in the arc-like filaments FA and FB displayed in Fig. \ref{fig:rcw108_boxes}.}
\label{fig:core_location}
\end{figure*}

{Figure~\ref{fig:core_location} shows the location of the 14 continuum cores.}
We find that the majority of the most massive cores are located along the filaments and around the high-mass stars shown in 
Fig. \ref{fig:rcw108_finding_chart} and
Fig. \ref{fig:chandra_fc_rcw108}.

\subsection{Feedback from IRS~29 on embedded cores in the arc-like filaments FA and FB}
\label{sec:results_yso_rcw108}

Three of the four most massive cores (22, 25, 29) are located along the arc-like filaments FA and FB (Fig. \ref{fig:rcw108_boxes}). To keep the interpretation conservative, we restrict the feedback-associated sample to these two filaments{, which contain the cores 35 (A-2) and 36 (A-1) in FA, and 22 (B-5), 23 (B-4), 25 (B-3), 27 (B-2), and 29 (B-1) in FB (Table~\ref{tab:FA_FB_summary}).}

\begin{table*}[htpb]
\centering
\caption{Properties of the conservative feedback-associated sample of cores in the arc-like filaments FA and FB.}
\label{tab:FA_FB_summary}
\setlength{\tabcolsep}{3pt}
\begin{tabular}{lcccccccc}
\hline\hline
Filament & IDs & $N$ & $d_{\rm proj}$ & $M_{20\,{\rm K}}$ & $M$ (MC) & $\tilde{\alpha}_{\rm vir}$ & $N(\alpha_{\rm vir}<2)$ & $N(P>0.84)$ \\
        &     &     & (pc) & ($M_\odot$) & ($M_\odot$) & & & \\
\hline
FA & 35 (A-2), 36 (A-1) & 2 & 0.066--0.067 & 7.36 & $6.9^{+3.3}_{-2.3}$ & $1.57^{+0.84}_{-0.55}$ & 1/2 & 1/2 \\
FB & 22 (B-5), 23 (B-4), 25 (B-3), 27 (B-2), 29 (B-1) & 5 & 0.100--0.145 & 34.21 & $32.5^{+14.8}_{-10.4}$ & $0.41^{+0.23}_{-0.15}$ & 5/5 & 5/5 \\
\hline
FA+FB & 35, 36, 22, 23, 25, 27, 29 & 7 & 0.066--0.145 & 41.57 & $39.5^{+17.7}_{-12.4}$ & $0.56^{+0.30}_{-0.20}$ & 6/7 & 6/7 \\
\hline
\end{tabular}
\tablefoot{{$d_{\rm proj}$ is the projected distance from IRS~29, and $\tilde{\alpha}_{\rm vir}$ is the median over the filament.} The number of cores that fulfill the virial threshold is indicated with $N(\alpha_{\rm vir}<2)$. {$N(P>0.84)$ gives the number of cores with a probability $P(\alpha_{\rm vir}<2)>0.84$ (Table~\ref{tab:continuum_cores_cmf_virial}).}}
\end{table*}

{The combined mass of the seven cores is $39.5^{+17.7}_{-12.4}\,M_\odot$.} The virial analysis shows that {six out of seven} FA+FB cores have \(\alpha_{\rm vir}<2\). {The exception is core 35 (A-2) in FA with $\alpha_{\rm vir}=2.51$.} The median virial parameter is $\tilde{\alpha}_{\rm vir}=1.57^{+0.84}_{-0.55}$ for FA and $\tilde{\alpha}_{\rm vir}=0.41^{+0.23}_{-0.15}$ for FB. {Hence,} the compact {cores} located along the arc-like filament FB are consistent with being gravitationally bound or prone to collapse under the adopted assumptions.

For $\rm Q_0=10^{48}\,{\rm s^{-1}}$, the FA cores at $\rm d_{\rm proj}\approx0.07$~pc are exposed to a projected ionizing photon flux of order $1.9\times10^{12}\,{\rm cm^{-2}\,s^{-1}}$. The FB cores, located at $\rm d_{\rm proj}\approx0.10-0.15$~pc, are exposed to fluxes of order $(4.0-8.4)\times10^{11}\,{\rm cm^{-2}\,s^{-1}}$.

{Since $\rm \Phi_i\propto Q_0$, additional ionizing sources of the local massive-star population would scale these fluxes linearly (cf. Sec.~\ref{sec:disc_irs29}).}

\subsection{H$_2$CO analysis of the thermal conditions in FA and FB}
\label{sec:H2CO_temp}

\begin{figure*}[h!]
	\centering
	\includegraphics[width=1.\textwidth]{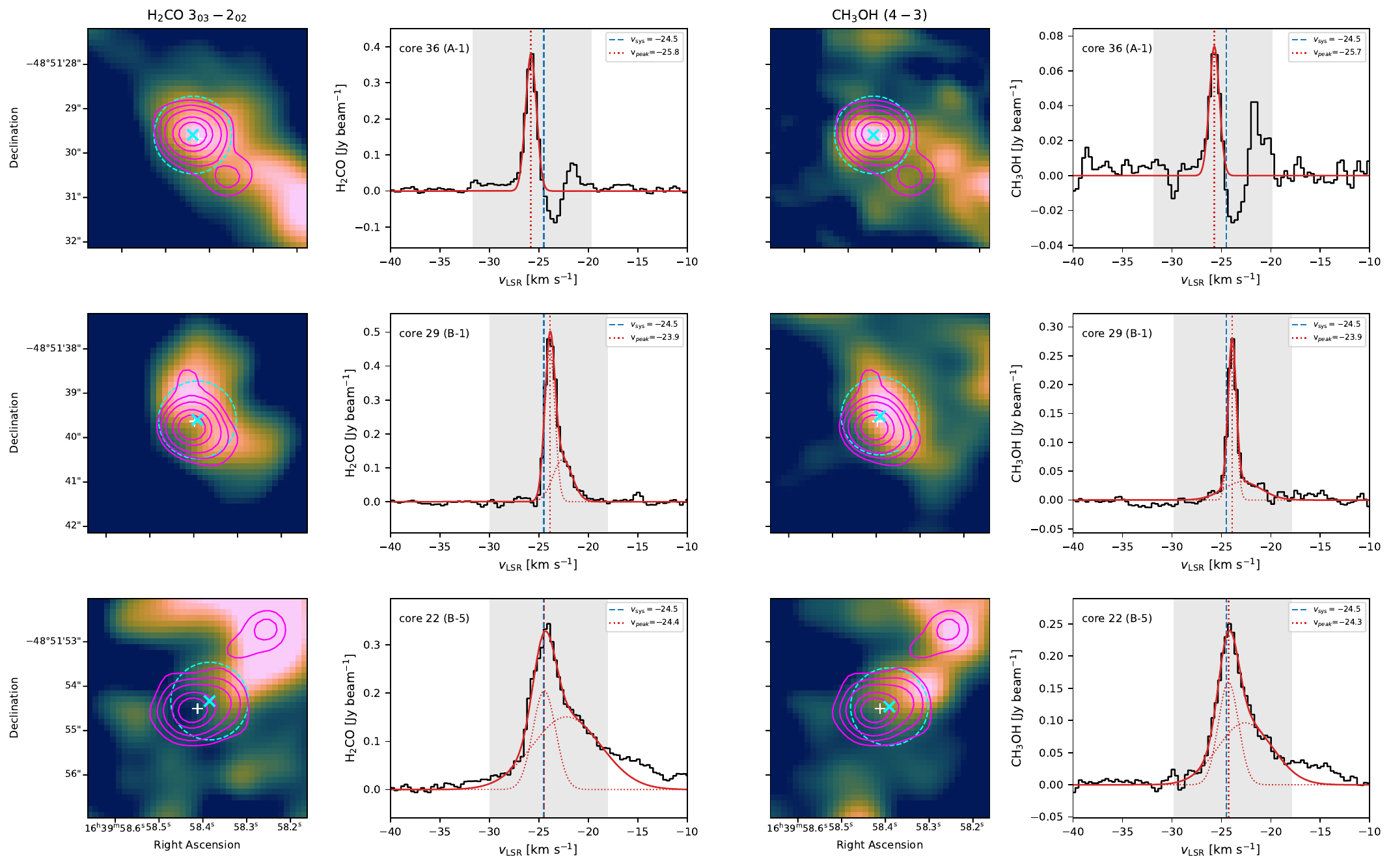}
	\caption{Molecular emission line counterparts located at the position of three continuum cores. On the left, we show H$_2$CO 3$_{03}$-2$_{02}$ {integrated-intensity maps and spectra}. On the right, the CH$_3$OH (4-3) line maps and their spectrum are shown. The pink contour lines represent the continuum core emission, the cyan dashed circle the aperture used for the spectrum. While core 36 (A-1) and core 29 (B-1) show compact H$_2$CO 3$_{03}$-2$_{02}$ and CH$_3$OH (4-3) emission at the continuum position, core 22 (B-5) shows an ambiguous picture.}
\label{fig:molecular_counterparts}
\end{figure*}

{The kinetic temperatures derived for the cores in FA and FB are listed in Table~\ref{tab:H2CO_FA_FB_grid}. In Fig.~\ref{fig:molecular_counterparts}, we show the H$_2$CO and CH$_3$OH molecular counterparts of the continuum cores analyzed in Sec. \ref{sec:Continuum_detection}. In this sample, the cores 36 (A-1) and 29 (B-1) show compact H$_2$CO emission at the position of the continuum core. The spectra of core 36 (A-1) show a narrow negative feature that is likely related to missing short-spacing (total-power) flux. For core 22 (B-5), the most massive core in our sample with M$_{\rm 20 K}\approx$13~M$_\odot$, both lines show a broad redshifted wing, and the line maps show emission peaks on opposite sides of the continuum core. A detailed analysis of this spectral structure will be presented in a subsequent paper.} 
{For the para-H$_2$CO line ratio, all six cores listed in Table~\ref{tab:H2CO_FA_FB_grid} reach an S/N of at least 4.5 in all three transitions, so that both line pairs independently satisfy the S/N threshold.}
\begin{table}[htbp!]
\centering
\caption{H$_2$CO temperature estimates for the continuum cores in the filaments FA and FB.}
\label{tab:H2CO_FA_FB_grid}
\setlength{\tabcolsep}{3pt}
\begin{tabular}{lccccc}
\hline\hline
Core & Filament &   T  & S/N & S/N & S/N \\
& &  [K] & 3$_{03}$-2$_{02}$  & 3$_{22}$-2$_{21}$ & 3$_{21}$-2$_{20}$ \\
\hline\hline
 &    &           & &                          &                   \\
36 (A-1) & FA &  $98.8^{+12.1}_{-6.6}$ & 34.3 & 9.5  & 11.4 \\
   &    &         &       &                              &            \\
35 (A-2) & FA &  $91.7^{+10.5}_{-3.2}$  & 15.8 & 13.4 & 13.7 \\
 &    &                           &            & &               \\
29 (B-1) & FB &    $113.9^{+2.9}_{-3.9}$ & 32.5 & 15.3 & 14.2\\
 &    &                           &            & &               \\
27 (B-2) & FB &    $81.4^{+3.4}_{-0.9}$ & 23.7 & 7.8  & 18.3 \\
   &    &                         &        &     &              \\
25 (B-3) & FB &   $>150$  & 14.3 & 13.3 & 12.4 \\
 &    &                           &            &  &              \\
22 (B-5) & FB &  $>150$ & 4.5  & 5.5  & 4.5  \\
 &    &                           &            &  &              \\
\hline
\end{tabular}
\tablefoot{
We use the para-H$_2$CO J=3-2 transitions 3$_{03}$-2$_{02}$, 3$_{22}$-2$_{21}$, and 3$_{21}$-2$_{20}$ and indicate the related signal-to-noise ratio of the investigated line. The uncertainties correspond to the 16-84 percentile range of the temperature estimates and therefore quantify the sensitivity to the extraction setup. We show the resulting temperatures T$_{\rm kin}$ as a function of the ratio (Eq. \ref{eq:kinetic_temperature1}) in Fig. \ref{fig:h2co_temperatures}. The H$_2$CO analysis of core 23 does not yield physically plausible temperature estimates and is excluded. {For core 22 and core 25, we find temperatures above the threshold of 150 K that distinguishes between valid and unreasonable model representations \citep{Magnum1993, Tang2017a}.}}
\end{table}
\begin{figure}[htpb]
	\centering
	\includegraphics[width=.5\textwidth]{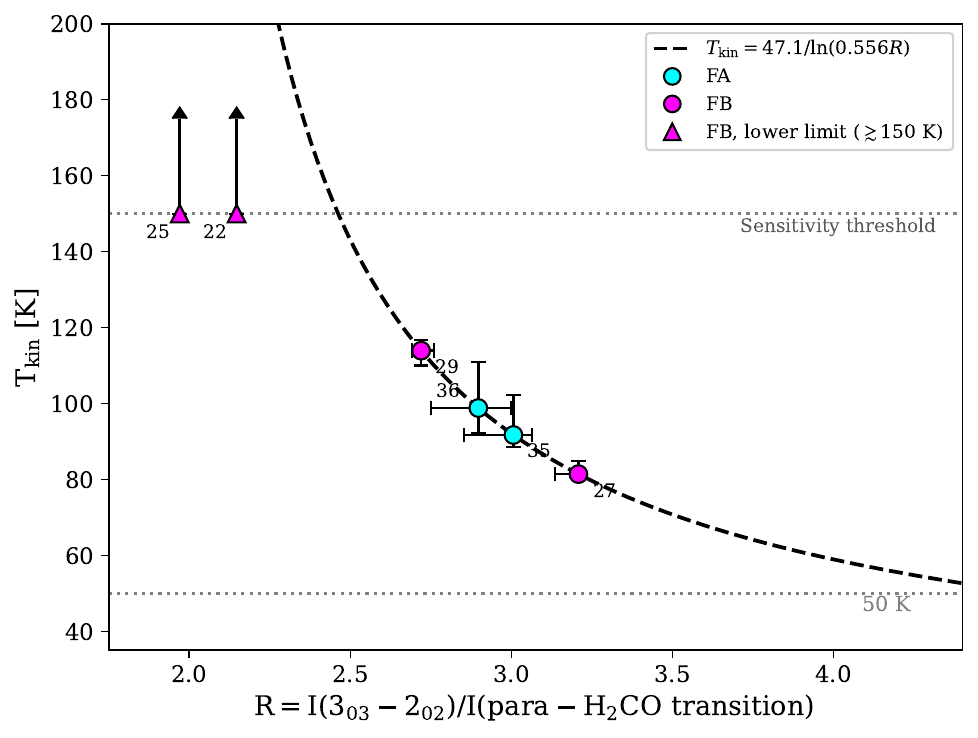}
	\caption{Temperature estimates for {cores 36, 35, 29, 27, 25, and 22} embedded in FA and FB using the para-H$_2$CO J=3-2 transition lines. The uncertainties of the parameter space are indicated. {For cores 22 and 25, the derived temperatures exceed the validity limit of 150~K and are shown as lower limits.} All cores can be classified as warm above the adopted temperature threshold of 50 K \citep{Tang2017a, Tang2017b}.}
\label{fig:h2co_temperatures}
\end{figure}
The temperature estimates indicate that the H$_2$CO emission associated with cores~{36, 35, 29, and 27} is warm, {averaging around 100~K and thus well} above the \(50\)~K threshold \citep{Tang2017a, Tang2017b}. 

{For cores 22 (B-5) and 25 (B-3), the line ratios are close to the validity limit $0.556R=1$, and the derived temperatures exceed 150~K, above which the analytic approximation is not reliable \citep{Magnum1993, Tang2017a}. We therefore only conclude that the H$_2$CO emission of these two cores is warm ($T_{\rm kin}\gtrsim150$~K). For core 22, the line ratios may additionally be affected by outflow-related or shock-heated gas (Fig.~\ref{fig:molecular_counterparts}). For core 23 (B-4), we do not obtain a physically plausible estimate.}

\section{Discussion}
\label{sec:discussion}

\subsection{Are FA and FB feedback-shaped filamentary structures?}
\label{sec:disc_fafb}

{Many of the detected cores are arranged along a previously unknown arc-like filamentary structure surrounding IRS\,29 and the nearby massive stars.}

{Such morphologies are frequently found around \HII\ regions, where pre- and protostellar cores are typically located in dense shells or fragmented filaments surrounding the ionized gas \citep{Deharveng2010, Zavagno2010}, comparable to the RCW~108 aggregate \citep{Comeron2005}.}

The {origin of the} arc-like filamentary structure in RCW~108 can have several possible explanations. Feedback from IRS~29, including stellar winds and ionizing radiation may have initially swept up the surrounding gas \citep{Elmegreen1977}. The stellar feedback in RCW\,108 could have produced the observed ring-like structure comparable to those reported for many \HII\ region bubbles, such as RCW~120 \citep{Luisi2021,Kabannovic2022} or RCW~79 \citep{Zavagno2010}. Stellar feedback potentially fragmented this ring (see \citealt{Bonne2023, Bonne2026} for observational support based on the ionized-carbon \CII 158 $\mu$m line), producing shattered fragments that retain an elongated arc-like morphology. This scenario is supported by the fact that RCW~108 is a known \HII\ region \citep{Comeron2005}, and that the area surrounding the central O9 star is devoid of molecular line and 1.3 mm emission (Fig.~\ref{fig:rcw108_finding_chart}). On the other hand, the eastern part of the region does not show arc-like structures, and the southern part exhibits a more filamentary morphology. It is therefore also possible that RCW~108 originated as a hub-filament system (HFS), within which the cluster formed, and that stellar feedback subsequently reshaped this pre-existing structure \citep{Myers2009, Peretto2014, Kumar2020}. The extent to which core formation is driven by stellar feedback is unclear.

{We find that the} cores in FB are approximately regularly spaced (Table \ref{tab:spaced_fragemention}) with an average distance of 7800 AU (0.038 pc), suggesting a quasi-periodic spaced fragmentation \citep{Beuther2015, Zhang2020, Li2024}. This widely reported phenomenon across both low-mass and high-mass star-forming filaments is usually interpreted in the context of the gravitational (or magnetohydrodynamic) instability of an infinite fluid cylinder \citep{Inutsuka1992}. Reported inter-core spacings span roughly 0.02–5 pc, depending on the scale and mass of the filament. Small values (0.03-0.1 pc) are found, for example, for NGC6334 \citep{Shimajiri2019} or G35.20-0.74  \citep{Sanchez-Monge2014}, while medium values (0.2-0.4 pc) were detected in IRDC 18223 by \cite{Beuther2015}. {Our reported average spacing falls into the range of these small values.} However, the classical cylindrical-fragmentation prediction is challenged by some observations \citep{Henshaw2016}, and it is proposed instead that the observed continuum or line-emission peaks, i.e., cores, trace a complex network of independent, gravitationally interacting sub-filaments rather than a single coherent cylinder fragmenting at one characteristic wavelength.

While we find signs of a spaced-fragmentation scenario, projection effects and the limited number of cores {do not support} a {single} interpretation.

\subsection{Star-forming potential of the embedded core-sample}
\label{sec:disc_sf}

{The cores in FB (Fig.~\ref{fig:rcw108_boxes}, right) and A-1 (core 36) in FA are approximately Gaussian-like. However, B-1 (core 29), B-3 (core 25), B-4/B-5 (core 23/22), and A-1 show hints of additional compact emission, which could indicate sub-fragmentation \citep{Goodwin2007, Bontemps2010}, similar to the protostellar condensations discussed by \cite{Andre2007}, but also blending or residual filamentary emission. Since typical fragmentation scales found by ALMAGAL are of the order of 10$^3$~AU \citep{Coletta2025, Molinari2025, Wallace2026}, comparable to our beam size (Table~\ref{tab:h2co_ch2oh_lines}), we cannot distinguish between these possibilities.}

{With {radii of $\leq${2830}~AU}, the four most massive cores ({4, 22, 25, 29}) are comparable to the compact fragments identified by ALMA-IMF and ALMAGAL \citep{Louvet2024, Coletta2025}. Their extended envelopes are likely interferometrically filtered out in the 12m continuum data, consistent with higher fluxes recovered in the 12m+7m maps (see Sec.~\ref{sec:limitation}).}

{Cores with $\alpha_{\rm vir}\geq2$ are not bound by gravity alone and, without an additional confining term such as external pressure from the ambient medium, would disperse on their crossing time $t_{\rm cross}=R/\sigma_v$ \citep{Bertoldi1992}. For the {5} cores with $\alpha_{\rm vir}\geq2$ ({35, 37, 30, 0, 24}), we find $t_{\rm cross}\approx({1.2}-{2.6})\times10^4$~yr, where the shortest crossing time belongs to core 24 with the largest virial parameter of $\alpha_{\rm vir}=5.15$. This is short compared to the cluster age of $\sim$1~Myr, so that these cores are either transient structures or confined by external pressure.}

{Studies of cloud ensembles and clumps, including the original work of \cite{Larson1981}, found either a power-law relation $\sigma_v\propto R^{0.2-0.5}$ or no correlation (see \citealt{Schneider2004, Schneider2011} for a compilation of literature values). Exponents between 0.33 and 0.5 are explained in the context of turbulence that is injected at large scales and cascades down to the smallest scales, while turbulence can also be driven on small scales by internal sources such as stellar winds and outflows. The supersonic, but size-independent C$^{18}$O line widths of the cores in RCW~108 (Sec.~\ref{sec:virial_analysis}) indicate that the non-thermal motions on these scales are not set by a turbulent cascade, but rather by local processes such as infall, outflows, or feedback-driven motions. However, the small sample, the narrow range of radii, and the spectral resolution ($\sigma_{\rm res}=0.566$~km\,s$^{-1}$) do not allow a firm conclusion. Despite these broad lines, most cores have $\alpha_{\rm vir}<2$ due to their high masses within small radii.}

{For core 23 (B-4), the C$^{18}$O line yields $\sigma_v={0.65}$~km\,s$^{-1}$ and an upper limit $\alpha_{\rm vir}<2$ (Table~\ref{tab:continuum_cores_cmf_virial}), whereas its $^{13}$CO line is considerably broader ($\sigma_v={1.06}$~km\,s$^{-1}$), which increases $\alpha_{\rm vir}$ by a factor of $\approx$3. The difference can be explained by the higher optical depth of $^{13}$CO and by blending. Since core 23 is unresolved and located next to the brightest core, 22 (B-5), the $^{13}$CO spectrum likely includes more extended, kinematically broader gas.} 

{The elevated H$_2$CO temperatures (Sec.~\ref{sec:H2CO_temp}) suggest that the molecular gas in parts of FA and FB is affected by local heating. Such heating may be related to embedded star formation activity, external radiative feedback from the central O9 star, or a combination of both. With our current data set, a specific heating mechanism remains ambiguous. We note that the adopted dust temperature of $T_{\rm dust}=20$~K refers to the continuum-emitting dust, whereas the para-H$_2$CO line ratios trace the kinetic temperature of warm, excited gas of about 100~K. The two temperatures therefore probe different components and are not directly comparable.}

{Furthermore, the position of core~36 (A-1) coincides with the candidate protostellar source CS-109 identified by \cite{Comeron2007}. Its H$_2$CO temperature of about 100~K is consistent with a locally heated, embedded protostar which supports ongoing star formation as proposed by \cite{Yamaguchi1999}.}

{The seven FA+FB cores (Sec.~\ref{sec:results_yso_rcw108}) therefore constitute a compact reservoir of dense gas that may contribute to future star formation in the feedback-affected environment of IRS~29.}

\subsection{Feedback-driven signatures from IRS~29}
\label{sec:disc_irs29}

{The projected ionizing photon fluxes from IRS~29 at the positions of the FA and FB cores are high compared to values reported for classical bright-rimmed clouds.} For example, \cite{Morgan2004} derive an ionizing flux of \(7.5\times10^8\,{\rm cm^{-2}\,s^{-1}}\) for the rim SFO~5 as part of the Sugitani, Fukui and Ogura (SFO) catalog \citep{Sugitani1991, Sugitani1994}, while resolved observations by \cite{Lefloch1997} give \(4.8\times10^9\,{\rm cm^{-2}\,s^{-1}}\) for the same object. 

The projected fluxes estimated for the embedded cores in FA and FB are therefore up to three orders of magnitude larger.

However, the exact local flux remains uncertain due to both the three-dimensional geometry and the unresolved contribution of nearby massive stars. From the works of \cite{Comeron2005} and \cite{Wolk2008}, IRS~29 is located within a compact group of sources with projected separations between 1500-3000 AU (Fig. \ref{fig:chandra_fc_rcw108}). As mentioned by \cite{Wolk2008}, the full aggregate of O and B stars appears responsible for the ionization of the diffuse emission in RCW~108. {In addition, \cite{Straw1987} reported a second O star in RCW~108, which has not yet been further investigated \citep{Comeron2007, Wolk2008}.} 

{Since Eq. \ref{eq:photon_flux} does not account for absorption by neutral hydrogen (through ionization) and dust \citep{Draine2011}, the incident flux on the cores might be lower, despite the contributions by additional stellar sources. Assuming a uniform medium, the flux decreases as 
\begin{equation}
    \rm \Phi(r)\,=\,Q_0/(4\pi r^2)\,[1-(r/R_{\rm S})^3]
\end{equation}
where $R_{\rm S} = (3Q_0/4\pi n_{\rm e}^2\alpha_{\rm B})^{1/3}$ is the Str\"omgren radius. For $Q_0 = 10^{48}$\,s$^{-1}$, this gives $R_{\rm S} \approx 0.3$\,pc for $n_{\rm e} = 10^3$\,cm$^{-3}$, i.e., a reduction of $\lesssim 10\%$ at the projected distances of FA and FB, but $R_{\rm S} \approx 0.07$\,pc for $n_{\rm e} = 10^4$\,cm$^{-3}$, which is comparable to the projected distance of FA. 
While a detailed analysis of the ambient medium is beyond the scope of this work, the incident photon flux might already be absorbed by the surface layer of the molecular filaments FA and FB, possibly resulting in the observed arc-like shapes. In this picture, the cores are shielded from the incident photon flux. Consequently, $\Phi_i$ estimated with Eq. \ref{eq:photon_flux} should be treated as an upper limit of the incident flux produced by IRS 29 on the surface of the molecular filaments of FA and FB.}

{On the basis of these results and in comparison with other works on feedback-associated star formation environments \citep{Lefloch1997, Morgan2004}, we find that the combination of the arc-like morphology, the proximity to IRS~29, the presence of {C$^{18}$O emission at the positions} of the continuum cores, and the low virial parameters is consistent with feedback-associated shaping and compression of dense molecular gas in FA and FB.}

\subsection{Implications for the stellar population}
\label{sec:future_population}

\cite{Comeron2007} proposed a top-heavy IMF for RCW~108, which provides context for interpreting the compact core population analyzed in this work. Based on near-infrared observations, \cite{Comeron2005} and \cite{Comeron2007} found that the embedded cluster RCW~108 contains fewer solar-type and low-mass stars than expected from an extrapolation of the bright end of the luminosity function. They suggested that external triggering may have contributed to an IMF that lacks the presence of low-mass stars.

{The derived CMF slope of $\alpha_{\rm CMF}={1.79}^{+{0.26}}_{-{0.16}}$ is shallower than the Salpeter value of 2.35 \citep{Salpeter1955} for the IMF, derived for stars in the solar neighborhood over the mass range roughly from 0.4 to 10 M$_\odot$. However, the precision allowed by this {14}-core sample does not allow us to confirm or reject a moderately top-heavy distribution as suggested for the region \citep{Comeron2007}.\newline 
We note that the} number of cores for the CMF analysis is comparable to works of, e.g., \cite{Beuther2004} and \cite{Zhang2015}. {In contrast, recent ALMA surveys base their estimated CMF on tens to several hundreds of cores \citep{Pouteau2022, Louvet2024, Traficante2026}.}

{While the core sample is quantitatively limited, the uncertainties of the CMF analysis (Appendix \ref{sec:uncertainties}) include core to core variations comparable to the methodology used in \cite{Traficante2026}. The authors of Traficante et al. find that a single dust temperature instead of core to core variations can significantly bias the slope of the CMF.}

{The comparison with previous studies of RCW~108 \citep{Comeron2005, Comeron2007, Wolk2008} therefore remains qualitative, and the present ALMA data alone do not establish a one-to-one mapping between the CMF and the IMF derived by \cite{Comeron2007}.}

{The massive stars in Fig.~\ref{fig:chandra_fc_rcw108} are not centrally concentrated with respect to the molecular cloud (Sec.~\ref{sec:stars}), which may point to a complex formation and dynamical history \citep{Zwart2010}. Variations of the CMF with the evolution of the star-forming environment \citep{Coletta2025} may be related to the dynamical impact of massive stars on their surroundings \citep{Sana2012, Offner2023, Bordier2024, Frost2025}. Testing the contribution of cluster collapse, mass segregation, or triggered star formation, and thus to what extent IRS~29 may have shaped the cluster, requires further observational and theoretical work.}

{The} compact {cores} may contribute to future star formation, but their final stellar content depends on the star formation efficiency, further fragmentation, and their subsequent dynamical evolution. The present observations do not establish whether they will form a distinct second generation of cluster members or whether they will modify the low-mass end of the IMF suggested by \cite{Comeron2007}. Given the young cluster age of approximately 1~Myr, stellar feedback and gas dispersal may influence the future dynamical evolution of RCW~108 \citep{Pfalzner2013}.

\subsection{Caveats of the analysis}
\label{sec:limitation}

{We discuss the impact of the choice between the three data sets (Table~\ref{tab:comp_data}, Sec.~\ref{subsec:alma}).}

{Interferometric filtering of extended emission produces negative bowls in the 12m archive data and decreases the flux of individual cores \citep{Sanhueza2019}, while additionally recovered extended emission increases the flux and thus the masses.}

{We have compared the impact of the three data sets on the 14 cores of our CMF sample (Fig.~\ref{fig:cmf_continuum}). Since the core sample is defined in the 12m archive map, we use this map as the reference. A dendrogram extraction with the same parameters recovers all 14 cores as individual leaves in the re-imaged 12m map. In the 12m+7m map, with its larger beam and about twice the noise, 9 of the 14 cores are recovered as individual leaves, three (8, 30, 35) are merged with a neighboring core, and two (0, 24) fall below the thresholds. Within the same apertures, the masses increase systematically, and the total mass of the 33 cores rises from 72.8~$M_\odot$ to 91.5~$M_\odot$ (re-imaged 12m) and 108.7~$M_\odot$ (12m+7m) for $T_{\rm dust}=20$~K. The increase is smallest for the compact, bright cores ($\approx$3\% for cores 22 and 23) and largest for the more extended ones, suggesting that the recovered extended emission is the main driver. Despite these differences, the maximum-likelihood CMF slope remains unchanged (1.79, 1.78, and 1.76), and the virial parameters decrease, so that our results based on the 12m archive map are conservative.}

{Missing short spacings also affect the line widths. In the 12m-archive data, 11 of the 14 cores have resolution-limited lines, more than half of them with nonphysical widths, including five of the seven cores in FA and FB. In the 12m+7m data, only two of the 14 cores (core 36 in FA and core 30) are resolution limited, which is why the virial analysis is based on these data.}

{We further tested the impact of the negative bowls in the $^{13}$CO data (Fig.~\ref{fig:rcw108_finding_chart}) by comparing them with single-dish SEST observations of the same region \citep{Comeron2005}. The column density derived from the 12m-archive $^{13}$CO data is higher than that from SEST (Appendix~\ref{sec:density_profile}), which implies that spatial filtering does not remove the high-column-density components.}

{For completeness, we emphasize an additional limitation of the feedback-driven interpretation of RCW~108. As discussed by \cite{Dale2015} and demonstrated in the simulations of \cite{Dale2013}, spatial coincidence between young stellar objects, dense gas structures, ionized interfaces, pillars, or shell-like features does not uniquely establish triggered star formation. In particular, triggered and spontaneously formed objects can be spatially mixed, and geometrical or age-sequence arguments based on a single snapshot can be ambiguous.
Both collect-and-collapse and radiatively driven implosion provide plausible physical frameworks for interpreting the arc-like filaments FA and FB. However, the available data do not allow us to distinguish uniquely between these mechanisms, nor to exclude the possibility that some of the dense structures were pre-existing and subsequently exposed or compressed by the feedback from IRS~29 and the embedded massive-star population.}

\section{Conclusions}
\label{sec:conclusions}

We have presented an ALMA Band~6 study of compact dense-gas structures in the embedded cluster RCW~108{, combining 233~GHz continuum and molecular-line data with the Chandra X-ray view of the embedded massive stellar population. Our key results are:}
\begin{enumerate}
    \item Out of {33} identified compact continuum cores, {14} are above the adopted lower-mass limit of \({1.0}\,M_\odot\) with a corresponding CMF slope of \(\alpha_{\rm CMF}={1.79}^{+{0.26}}_{-{0.16}}\).
    \item The continuum core masses span \({0.16}-{12.72}\,M_\odot\) with \(M_{\rm total}={69.4}^{+{29.9}}_{-{21.3}}\,M_\odot\).
    \item {All {14} continuum cores above the mass limit show C$^{18}$O emission at their position.}
    \item {Nine of the 14 cores have \(\alpha_{\rm vir}<2\), consistent with being gravitationally bound or prone to collapse.}
    \item {The C$^{18}$O line widths are supersonic, independent of core size, and lie a factor of $\approx$2–4 above the Larson-relation.}
    \item {Six of the seven cores in the arc-like filaments FA and FB have \(\alpha_{\rm vir}<2\). For the majority of cores in FA and FB, we find kinetic temperatures above 50 K estimated from the para-H$_2$CO line ratios.}
    \item Assuming an O9V-equivalent ionizing photon rate, {the projected ionizing photon flux at FA and FB exceeds that of classical bright-rimmed clouds by up to three orders of magnitude, while the surface layers of the filaments could shield the embedded cores.}
\end{enumerate}
Taken together, the spatial coincidence between compact continuum cores, arc-like molecular gas, low virial parameters in FA and FB, and the projected ionizing radiation field from IRS~29 {is consistent with} feedback-associated compression of dense gas in RCW~108. {Future observations of further signatures of feedback-driven processes will help to address the remaining caveats of the analysis.}

\begin{acknowledgements}
F.P., L.L., E.B., L.G., N.Schm., M.M., R.S., N.S., S.D., P.S., S.P., and C.N. gratefully acknowledge the Collaborative Research Center 1601 funded by the Deutsche Forschungsgemeinschaft (DFG, German Research Foundation) -- SFB 1601 [sub-projects A3, B2, A2, A5] -- 500700252 and the Cluster of Excellence “Our Dynamic Universe” (DYNAVERSE). S.K. acknowledges support by the BMWI via DLR, project number 50OR2311. M.M. was supported for this research through a stipend from the International Max Planck Research School (IMPRS) for Astronomy and Astrophysics at the Universities of Bonn and Cologne. This research made use of astrodendro, a Python package to compute dendrograms of Astronomical data (\url{http://www.dendrograms.org/}). A.S-M.\ acknowledges support from the PID2023-146675NB grant funded by MCIN/AEI/10.13039/501100011033, and by the programme Unidad de Excelencia Mar\'{\i}a de Maeztu CEX2020-001058-M, as well as support from the RyC2021-032892-I grant funded by MCIN/AEI/10.13039/501100011033 and by the European Union `Next GenerationEU'/PRTR.
This paper makes use of the following ALMA data: ADS/JAO.ALMA\#2016.1.01182.S. ALMA is a partnership of ESO (representing its member states), NSF (USA) and NINS (Japan), together with NRC (Canada), NSTC and ASIAA (Taiwan), and KASI (Republic of Korea), in cooperation with the Republic of Chile. The Joint ALMA Observatory is operated by ESO, AUI/NRAO and NAOJ. This research has made use of data obtained from the Chandra Data Archive provided by the Chandra X-ray Center (CXC). This work is based in part on observations made with the Spitzer Space Telescope, which was operated by the Jet Propulsion Laboratory, California Institute of Technology under a contract with NASA.

\end{acknowledgements}

\bibliographystyle{aa}
\bibliography{bib}

\appendix

\section{Uncertainties and error propagation}
\label{sec:uncertainties}

{We propagate all uncertainties with a Monte Carlo approach using $10^4$ realizations. The adopted uncertainties are summarized in Table~\ref{tab:uncertainty_budget}. We distinguish between statistical uncertainties, which are drawn independently for each core, and systematic uncertainties, which are common to all cores and are drawn once per realization.}

\begin{table}[htbp]
\centering
\scriptsize
\caption{Adopted uncertainties and their propagation.}
\label{tab:uncertainty_budget}
\setlength{\tabcolsep}{3pt}
\begin{tabular}{lp{3.3cm}ll}
\hline\hline
Source & Adopted value & Type & Affects \\
\hline
Flux noise & $\sigma_{\rm cont}/{\rm PB}_i\sqrt{N_{{\rm pix},i}/N_{\rm beam}}$, $N_{\rm beam}=30.3$ & statistical & $M$ \\
Flux calibration & 10\% (Gaussian) & systematic & $M$ \\
$T_{\rm dust}$ & 15-30~K (uniform) & per core & $M$, $\alpha_{\rm vir}$, CMF \\
$\kappa_{233\,{\rm GHz}}$ & factor 2 $=2\sigma$ (log-normal) & systematic & $M$, $\alpha_{\rm vir}$ \\
Gas-to-dust ratio & 100-110 (uniform) & systematic & $M$, $\alpha_{\rm vir}$ \\
$\sigma_{\rm obs}$ & fit covariance $\times1.41$ (Hanning) & statistical & $\alpha_{\rm vir}$ \\
$\sigma_v$ (RL) & upper limit, $\max(\sigma_{\rm obs},\sigma_{\rm res})$ & - & $\alpha_{\rm vir}$ \\
$R$ & dendrogram parameters $\pm1\sigma$ $\oplus$ pixel bootstrap & statistical & $\alpha_{\rm vir}$ \\
$Q_0$ & $\log Q_0=47.9$-48.1 & systematic & $\Phi$ \\
\hline
\end{tabular}
\tablefoot{Statistical uncertainties are drawn independently for each core, systematic uncertainties once per realization for all cores ($10^4$ realizations). The distance (1300~pc) enters as a scaling: $M\propto d^2$, $R\propto d$, $\alpha_{\rm vir}\propto d^{-1}$.}
\end{table}

{The statistical flux uncertainty of each core is $\sigma_{S,i}=\sigma_{{\rm cont},i}\sqrt{N_{{\rm pix},i}/N_{\rm beam}}$, where $N_{{\rm pix},i}$ is the number of pixels of the dendrogram leaf, $N_{\rm beam}={30.3}$ is the number of pixels per beam, and $\sigma_{{\rm cont},i}$ is the {noise of the primary-beam-corrected map at the core position ($\sigma_{\rm cont}$ divided by the mean primary-beam response over the leaf).} In addition, we adopt an absolute flux calibration uncertainty of 10\% for ALMA Band~6\footnote{ALMA Technical Handbook, \url{https://almascience.eso.org/documents-and-tools}}. The dust temperature is drawn from a uniform distribution between 15 and 30~K for each core individually, since the cores may be heated differently. For the dust opacity, \cite{Ossenkopf1994} estimate an uncertainty of about a factor of two{, which we treat as the $2\sigma$ width of a log-normal distribution}. The gas-to-dust ratio of 100 implicit in $\kappa_{233\,{\rm GHz}}$ may increase to $\approx$110 at the galactocentric distance of RCW~108 of 7~kpc \citep{Giannetti2017}.} {The observed line widths $\sigma_{\rm obs}$ are drawn from Gaussian distributions with the fit uncertainties. For RL emission lines, each draw is replaced by $\max(\sigma_{\rm obs},\sigma_{\rm res})$, so that the upper-limit character is preserved. The uncertainty of {$r_{\rm obs}$} combines the variation of the dendrogram leaf when the minimum intensity and contrast are changed by $\pm1\,\sigma_{\rm cont}$ with a bootstrap resampling of the leaf pixels, which is propagated to {$\delta r_{\rm dec}=r_{\rm obs}\,\delta r_{\rm obs}/r_{\rm dec}$}. Radii of unresolved cores are kept at their upper limit.}

{For each quantity, we report the median and the 16-84 percentile range of the resulting distribution, given separately for the statistical and the total uncertainty. For the virial analysis, we additionally provide the probability $P(\alpha_{\rm vir}<2)${, which is a lower limit for cores with upper limits on $\alpha_{\rm vir}$}.}

{To test the assumptions of optically thin emission and of the adopted temperature range, we compare the peak brightness temperature $T_{\rm B}$ of each core with the Planck-corrected temperature $J_\nu(T_{\rm dust})$. For a source that fills the beam, $T_{\rm B}=J_\nu(T_{\rm dust})\,(1-e^{-\tau})\leq J_\nu(T_{\rm dust})$. Hence, $T_{\rm B}$ provides a lower limit to $T_{\rm dust}$, and for a given $T_{\rm dust}$, the peak optical depth follows from $\tau=-\ln\left[1-T_{\rm B}/J_\nu(T_{\rm dust})\right]$. The brightest core, core 22, has a peak intensity of 0.197~Jy~beam$^{-1}$, corresponding to $T_{\rm B}\approx9.8$~K. This excludes dust temperatures below $\approx$14.7~K, consistent with the adopted lower limit of 15~K. For $T_{\rm dust}=20$~K, its peak optical depth is $\tau\approx1.08$, and its mass may be underestimated by up to a factor of $\tau/(1-e^{-\tau})\approx1.63$. For all other cores, we find $\tau\leq0.52$, with the largest value for core 29. The related emission is therefore at most moderately optically thick.}

\section{Core catalog}
\label{app:13CO_catalog}

Here, we list all {33} {cores} detected in the continuum. The catalog is defined from independent dendrogram leaves using a ${3}\sigma_{\rm cont}$ intensity threshold, a ${9}\sigma_{\rm cont}$ minimum contrast, and an area that matches the beam size of the ALMA observations. The masses listed here are computed for the dust temperature $T_{\rm dust}=20$~K and $\kappa_{233\,{\rm GHz}}=0.009\,{\rm cm^2\,g^{-1}}$ for gas and dust \citep{Ossenkopf1994}.

\begin{table}[htpb]
\centering
\setlength{\tabcolsep}{4pt}
\caption{Dust-continuum cores identified in RCW~108 at 233~GHz.}
\label{tab:continuum_cores}
\begin{tabular}{ccccccc}
\hline\hline
ID & R.A. & Dec & Flux & Mass & $r_{\rm obs}$ & $R$ \\
   & (deg) & (deg) & ($10^{-2}$ Jy) & (M$_\odot$) & (arcsec) & (arcsec) \\
\hline
0 & 250.0076 & -48.8740 & 4.82 & 1.74 & 1.02 & 1.88 \\
1 & 249.9928 & -48.8746 & 0.73 & 0.26 & 0.31 & 0.23 \\
2 & 250.0038 & -48.8742 & 2.19 & 0.79 & 0.50 & 0.79 \\
4 & 250.0032 & -48.8718 & 27.1 & 9.80 & 1.18 & 2.18 \\
6 & 250.0015 & -48.8704 & 1.88 & 0.68 & 0.47 & 0.72 \\
7 & 250.0237 & -48.8703 & 0.992 & 0.36 & 0.29 & $\lesssim$0.54 \\
8 & 250.0018 & -48.8694 & 4.12 & 1.49 & 0.59 & 0.98 \\
9 & 249.9997 & -48.8695 & 0.992 & 0.36 & 0.38 & 0.49 \\
10 & 250.0137 & -48.8693 & 0.618 & 0.22 & 0.24 & $\lesssim$0.46 \\
11 & 249.9981 & -48.8689 & 2.02 & 0.73 & 0.56 & 0.91 \\
12 & 249.9950 & -48.8686 & 0.876 & 0.32 & 0.48 & 0.73 \\
14 & 249.9948 & -48.8670 & 0.662 & 0.24 & 0.28 & $\lesssim$0.53 \\
18 & 249.9948 & -48.8665 & 1.45 & 0.53 & 0.42 & 0.59 \\
20 & 249.9946 & -48.8658 & 1.05 & 0.38 & 0.25 & $\lesssim$0.47 \\
22 & 249.9934 & -48.8651 & 35.2 & 12.72 & 0.40 & 0.53 \\
23 & 249.9928 & -48.8647 & 4.51 & 1.63 & 0.28 & $\lesssim$0.54 \\
24 & 249.9868 & -48.8643 & 4.78 & 1.73 & 0.90 & 1.63 \\
25 & 249.9916 & -48.8638 & 22 & 7.93 & 0.59 & 0.99 \\
26 & 250.0065 & -48.8635 & 0.616 & 0.22 & 0.24 & $\lesssim$0.45 \\
27 & 249.9924 & -48.8626 & 12.6 & 4.57 & 0.88 & 1.60 \\
28 & 249.9839 & -48.8616 & 0.432 & 0.16 & 0.22 & $\lesssim$0.42 \\
29 & 249.9937 & -48.8610 & 20.4 & 7.35 & 0.53 & 0.85 \\
30 & 249.9995 & -48.8598 & 6.81 & 2.46 & 1.30 & 2.42 \\
32 & 250.0046 & -48.8594 & 2.4 & 0.87 & 0.61 & 1.03 \\
33 & 250.0036 & -48.8593 & 6.33 & 2.29 & 0.74 & 1.31 \\
34 & 249.9955 & -48.8596 & 1.44 & 0.52 & 0.51 & 0.80 \\
35 & 249.9973 & -48.8590 & 8.86 & 3.20 & 1.10 & 2.03 \\
36 & 249.9997 & -48.8582 & 11.5 & 4.16 & 0.44 & 0.63 \\
37 & 250.0040 & -48.8571 & 7.01 & 2.53 & 1.27 & 2.36 \\
38 & 250.0006 & -48.8560 & 2.46 & 0.89 & 0.66 & 1.14 \\
39 & 250.0025 & -48.8553 & 2.35 & 0.85 & 0.46 & 0.69 \\
40 & 249.9895 & -48.8546 & 0.553 & 0.20 & 0.23 & $\lesssim$0.44 \\
41 & 250.0037 & -48.8502 & 1.87 & 0.68 & 0.51 & 0.81 \\
\hline
\end{tabular}
\tablefoot{Independent dendrogram leaves ($3\sigma_{\rm cont}$ threshold, $9\sigma_{\rm cont}$ minimum contrast, 28 pixels minimum area). Masses for $T_{\rm dust}=20$~K. $R=1.91\,r_{\rm dec}$ (Table~\ref{tab:continuum_cores_cmf}).}
\end{table}

From these {33} cores, a subset of {14} objects above the lower mass level of {1.0}~M$_{\odot}$ is created that are listed in Table \ref{tab:continuum_cores_cmf}.

\section{C$^{18}$O and $^{13}$CO spectra of all cores}
\label{app:conti_13co_pairs}

\begin{figure*}[htpb]
\centering \includegraphics[width=.9\textwidth]{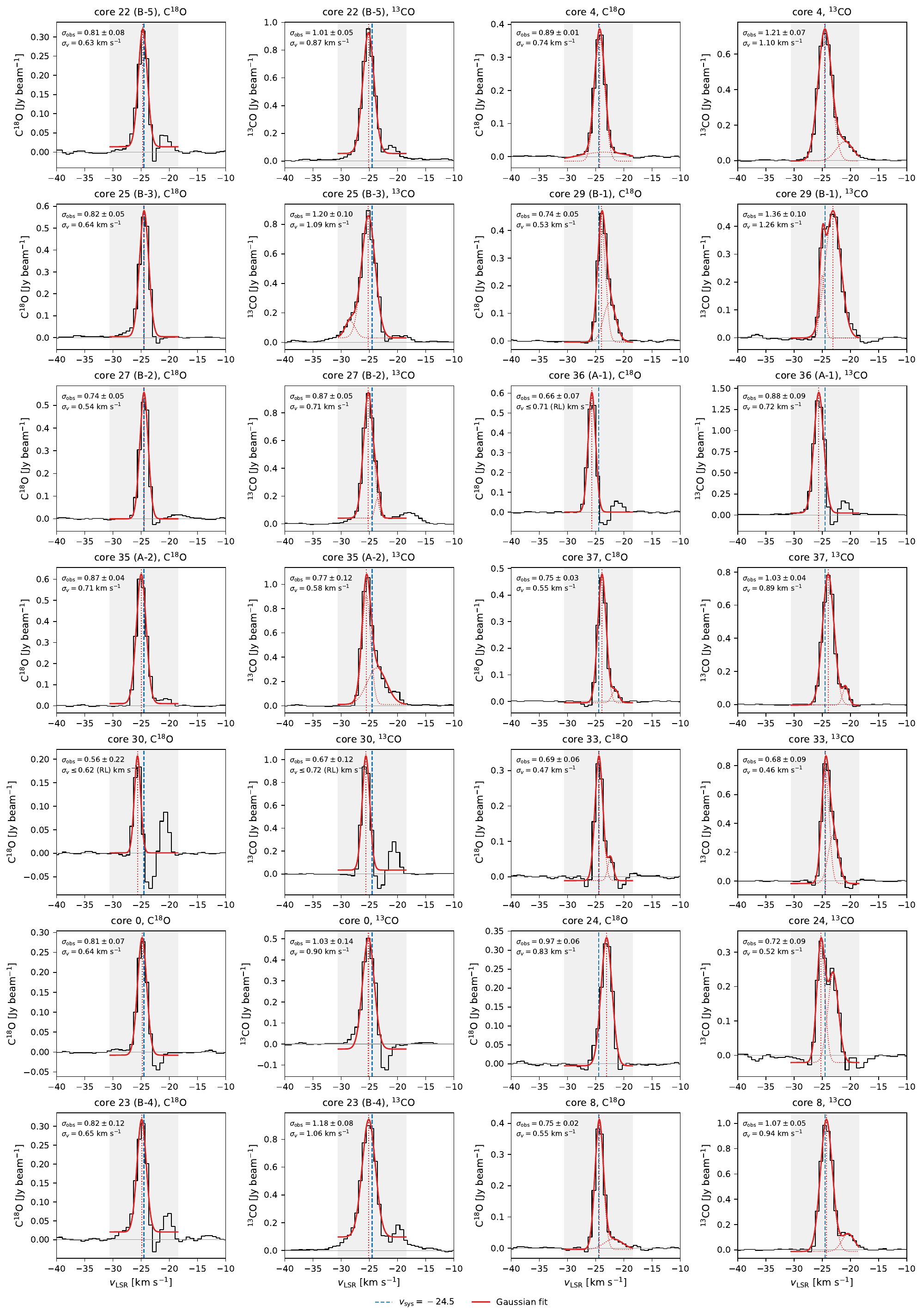}
\caption{{C$^{18}$O\,(2-1) and $^{13}$CO\,(2-1) spectra of all {14} continuum cores above ${1.0}\,M_\odot$, averaged over the continuum leaf masks (two cores per row, C$^{18}$O and $^{13}$CO side by side). Symbols and lines as in Fig.~\ref{fig:spectra_main}.}}
\label{fig:continuum_13co_pairs}
\end{figure*}

{Figure~\ref{fig:continuum_13co_pairs} shows the C$^{18}$O\,(2-1) and $^{13}$CO\,(2-1) spectra of all {14} cores above ${1.0}\,M_\odot$, averaged over the continuum leaf masks, together with the Gaussian fits used for the virial analysis (Sec.~\ref{sec:data_core_detection}). The results for $^{13}$CO and for {$R={1.177}\,r_{\rm dec}$} are used as a comparison (Sec.~\ref{sec:virial_analysis}). The $^{13}$CO emission is detected toward all cores, but its lines are {mostly} broader than those of C$^{18}$O. For one core, its line width is resolution-limited, compared to two for C$^{18}$O. Two-component fits are adopted for {5} C$^{18}$O and {9} $^{13}$CO spectra. We find for core 25 (B-3) that the second $^{13}$CO component is a broad blue-shifted feature that may be associated with outflowing gas.}

{Each spectrum is fitted within $v_{\rm sys}\pm6$~km\,s$^{-1}$ with a Gaussian and a constant baseline. Following \cite{Kass1995}, a second Gaussian component is admitted if it lowers the Bayesian information criterion by more than 2 and both components are constrained. In that case, the brighter component is adopted. A line is considered detected if its fitted peak exceeds five times the RMS of the line-free channels of the same spectrum ($|v-v_{\rm sys}|>8$~km\,s$^{-1}$). Since neighboring channels are correlated by the Hanning smoothing, the formal fit uncertainties are multiplied by $\sqrt{2}$. For lines whose observed width is close to the spectral resolution ($\sigma_{\rm obs}<1.2\,\sigma_{\rm res}$), the instrumental correction in Eq.~\ref{eq:sigma_v} is not reliable. These resolution-limited (RL) cores remain in the sample with upper limits on $\sigma_v$, $M_{\rm vir}$, and $\alpha_{\rm vir}$.}

\section{Consistency test of the $^{13}$CO column density}
\label{sec:density_profile}

Because many of the most massive continuum cores are located in the filaments of RCW~108 (Fig. \ref{fig:rcw108_finding_chart} and Fig. \ref{fig:core_location} ), we inspect the filaments FA and FB that are arranged in an arc-like feature shown in Fig. \ref{fig:ALMA_fc_12m_archive}. From this feature, we derive a peak flux density per beam for the $\rm ^{13}CO$ line of 0.39 Jy/beam. We use CASA and QFitsView \citep{Ott2013} to derive the peak flux.
The flux density translates to a corresponding brightness temperature of $T_{\rm Bt}\,=\,22.3\,$K. To estimate the excitation CO temperature $T_{\rm ex}$ we use the optically thick $^{12}$CO~(2-1) brightness temperatures  $T_{\rm Bt} = 30.1\,{\rm K}$ reported by \cite{Comeron2005}. The excitation temperature is calculated as follows:
\begin{equation}
T_{\mathrm{ex}} =\frac{T_0}{\ln\left(\frac{T_0}{T_{\mathrm{Bt}}}\left(1-e^{-\tau}\right)+1\right)}
\overset{\tau \gg 1}{\approx} \frac{T_0}{\ln\left(\frac{T_0}{T_{\mathrm{Bt}}}+1\right)}
\end{equation}
with $T_0 = h\nu/k_{\rm B}$. We obtain a CO excitation temperature of $T_{\rm ex} = 35.34\,{\rm K}$
for the dense filament. Since the $^{13}$CO line is only moderately optically thick, we determine the corresponding column density from the integrated brightness temperature in the optically thin limit \citep{Mangum2015} and apply an opacity correction below:
\begin{align}
    N_{\rm thin}({\rm ^{13}CO}) &= \frac{3h}{8\pi^3\mu_d^2J_uT_0}\left(\frac{k_{\rm B} T_{\rm ex}}{hB_0}+\frac{1}{3}\right)e^{\frac{T_u}{T_{\rm ex}}}\int T_{\rm Bt}\,{\rm d}v \nonumber \\
    &=1.17\times10^{13}(T_{\rm ex}+0.88)e^{\frac{15.87 K}{T_{\rm ex}}}\int T_{\rm Bt}\,{\rm d}v ~,
\end{align}
with $\mu_d$ as the dipole moment, the upper J-level $\rm J_u$, and $\rm B_0$ the rotational constant \citep{Kabannovic2022}. With the equivalent temperature of the upper energy J-level, we get
\begin{equation}
    T_u = \frac{E_u}{k_{\rm B}}  = \frac{h B_0 J_u(J_u+1)}{k_{\rm B}}~.
\end{equation}
{In order to convert the derived $^{13}$CO column density to the $^{12}$CO column density, it is necessary to multiply by the local carbon abundance ratio, $\alpha=62\pm 10$. The radial galactic carbon isotope abundance was derived by \cite{Milam2005} and the galactocentric distance of RCW~108 is 7 kpc \citep{Wolk2008, Peissker2022}.}
To take into account possible optical depth effects, we can use the optical depth from the observed ratio $^{12}$CO~(2-1) to $^{13}$CO~(2-1) \citep{Kabannovic2022}:
\begin{equation}
    \frac{T_{\rm Bt}({\rm ^{12}CO})}{T_{\rm Bt}({\rm ^{13}CO})}
    = \frac{1-e^{-\tau}}{1-e^{-\tau/\alpha}}
    \overset{\tau/\alpha \ll 1}{\approx} \frac{1-e^{-\tau}}{\tau} \alpha~,
\end{equation}

which result in an optical depth of $\rm \tau (^{12}CO) = 84$ and corresponding $\rm \tau (^{13}CO) = 1.35$.
This allows us to correct the previously derived column density:
\begin{equation}
N = N_{\rm thin}\frac{\tau}{1-e^{-\tau}}
\end{equation}
where we integrate over a range of [-30, 0]\,km/s, resulting in a $\rm ^{13}CO$ column density of $\rm N_{\rm ^{13}CO}=4.7\times10^{17}\,cm^{-2}$.
Using an assumed abundance ratio of $\rm ^{13}CO$ to $\rm ^{12}CO$ of 62 and a $\rm ^{12}CO/H_2$ conversion factor of $8.5\times10^{-5}$ from \cite{Tielens2010} and \cite{Kabannovic2022}, we derive the total molecular hydrogen column density of  $\rm N_{\rm H_2 }=3.4\times10^{23}\,cm^{-2}$.
We note that the estimated hydrogen column density $\rm N_{\rm H_2 }$ in this work is about a factor of five higher compared with single-dish observations executed with SEST \citep{Comeron2005}. Part of this offset reflects the different conversion from $\rm ^{13}CO$ to $\rm H_2$ adopted in the two studies. For $\rm N(^{13}CO)$ alone, our value is about three times higher. The higher column density is consistent with the interpretation that the compact 1.3 mm cores identified in this work trace the high-column-density component of the molecular material located in FA and FB (Fig. \ref{fig:core_location}), which was previously detected with SEST \citep{Comeron2005} but neither spatially resolved nor separated from the averaged extended emission.

\section{H$_2$CO temperature analysis}
\label{app:h2co_details}

{We extract the H$_2$CO spectra at the continuum-selected core positions and fit the line profiles with Gaussian profiles to determine the para-H$_2$CO ratios within a grid of extraction parameters (Table~\ref{tab:H2CO_parameter_grid}). This grid includes different aperture sizes, velocity windows, local baseline treatments, and H$_2$CO-based velocity centers. In total, the parameter space includes 420 settings per core (2940 in total). Since offsets between the 1.3~mm continuum and H$_2$CO may be explained by different properties of formaldehyde, we allow for minor deviations between the core position and the molecular gas line. The resulting line velocities are compared to the systemic velocity of RCW~108 of v$_{\rm LSR}$= -24.5 km/s to avoid unwanted confusion. Due to sporadic emission, we use the robust Gaussian-based line ratios as our reference temperature estimate.}
\begin{table}[h!]
\centering
\caption{Parameter space explored for the H$_2$CO temperature stability analysis.}
\label{tab:H2CO_parameter_grid}
\begin{tabular}{ll}
\hline\hline
Parameter & Settings  \\
\hline
Aperture radius & $0.5$, $0.75$, $1.0$, $1.25\,\theta_{\rm beam}$ \\
Velocity half-width & $0.5$, $0.7$, $1.0$, $1.5$, $2.0~\rm km/s$ \\
Velocity center & peak$_{303}$, gauss$_{303}$, systemic  \\
Baseline subtraction & off, on \\
Baseline inner gap & $0.5$, $1.0~\rm km/s$ \\
Baseline window width & $2.0$, $3.0$, $4.0~\rm km/s$ \\
RMS estimate & spectrum-based rms \\
Line-ratio validity & $\rm 0.556\,R > 1$ \\
Signal-to-noise threshold & $\rm (S/N) \geq 2.5$ \\
\hline
\end{tabular}
\tablefoot{The parameter grid is evaluated independently for each continuum-selected core. The baseline parameters are only applied when baseline subtraction is enabled. The terms peak$_{303}$ and gauss$_{303}$ refer to the velocity centers determined from the transition of the para-H$_2$CO ratio. The systemic option uses a fixed reference velocity and is included as a control case.}
\end{table}
{For each grid configuration, the 3$_{03}$-2$_{02}$ transition is used as the reference line for the Gaussian decomposition. We first determine the line center and width of the 3$_{03}$-2$_{02}$ emission and then fit the \(3_{22}-2_{21}\) and \(3_{21}-2_{20}\) transitions using the same velocity structure. This approach reduces the impact of individual noisy channels and provides a consistent estimate of the line ratios $\rm R_{322} = I(3_{03}-2_{02})/I(3_{22}-2_{21})$ and $\rm R_{321} = I(3_{03}-2_{02})/I(3_{21}-2_{20})$ for each extraction setup, which we treat equally \citep{Tang2017a, Tang2017b}. For each grid point, we estimate the local rms from the extracted spectrum. A temperature estimate is accepted if one of the fitted line pairs satisfies $({\rm S/N})\geq2.5$ and the corresponding ratio fulfills $0.556R>1$, so that Eq.~\ref{eq:kinetic_temperature1} returns a physical value. The quoted temperatures in Table~\ref{tab:H2CO_FA_FB_grid} correspond to the median of the valid Gaussian solutions across the parameter grid, while the uncertainties represent the 16-84 percentile range. These uncertainties therefore quantify the sensitivity to the extraction setup rather than formal statistical errors.}

\end{document}